\documentclass[10pt,eqsecnum
]{article}
\usepackage{amsmath,amsfonts,amssymb,latexsym,theorem,mathrsfs,latexsym,cite,setspace,
comment,color,multirow,ulem,graphicx}
\pdfoutput=1

\newcommand{\ma}[1]{\mbox{$\mathcal{#1}$}}
\newcommand{\mas}[1]{\mbox{$\mathscr{#1}$}}

\newcommand{\D}{{\rm d}}
\newcommand{\ti}{\tilde}

\begin{document}

\begin{titlepage}

\begin{flushright}
{
\today
}
\end{flushright}
\vspace{1cm}

\begin{center}
{\LARGE \bf
\begin{spacing}{1}
Robinson-Trautman solutions with scalar hair\\
 and Ricci flow
\end{spacing}
}
\end{center}
\vspace{.5cm}

\begin{center}
{\large \bf
Masato Nozawa${}^1$ and Takashi Torii${}^2$
} \\

\vskip 1cm
{\it
${}^1$General Education, Faculty of Engineering, 
Osaka Institute of Technology, Osaka City, Osaka 535-8585, Japan \\
${}^2$Department of System Design, Osaka Institute of Technology, Osaka City, Osaka 530-8568, Japan
}\\
{\texttt{masato.nozawa@oit.ac.jp, 
takashi.torii@oit.ac.jp}}

\end{center}

\vspace{.5cm}

\begin{abstract}
The vacuum Robinson-Trautman solution admits a shear-free and twist-free null geodesic congruence with a nonvanishing expansion. We perform a comprehensive classification of solutions exhibiting this property in Einstein's gravity  with a massless scalar field, assuming that the solution belongs at least to Petrov-type II and some of the components of Ricci tensor identically vanish. We find that these solutions can be grouped into three distinct classes: (I-a) a natural extension of the Robinson-Trautman family incorporating a scalar hair satisfying the time derivative of the Ricci flow equation, (I-b) a novel non-asymptotically flat solution characterized by two functions satisfying Perelman's pair of the Ricci flow equations, and (II) a dynamical solution possessing ${\rm SO}(3)$, ${\rm ISO}(2)$ or ${\rm SO}(1,2)$ symmetry. We provide a complete list of all explicit solutions falling into Petrov type D for classes (I-a) and (I-b). Moreover, leveraging the massless solution in class (I-a), we derive the neutral Robinson-Trautman solution to the  ${\cal N}=2$ gauged supergravity with the prepotential $F(X) =-iX^0X^1$. By flipping the sign of the kinetic term of the scalar field, the Petrov-D class (I-a) solution leads to a time-dependent wormhole with an instantaneous spacetime singularity. Although the general solution is unavailable for class (II), we find a new  dynamical solution with spherical symmetry from the AdS-Roberts solution via AdS/Ricci-flat correspondence. 
\end{abstract}

\vspace{.5cm}

\setcounter{footnote}{0}

\end{titlepage}

\setcounter{tocdepth}{2}
\tableofcontents

\newpage

\section{Introduction}

The construction of exact solutions to Einstein's equations has been a convoluted and challenging
problem in general relativity. In pursuit of  this objective, the study of ray optics has played a prominent role. 
According to the Goldberg-Sachs theorem, 
the vacuum spacetime possessing a shear-free null geodesic congruence is algebraically special \cite{GS}, i.e., 
there exist repeated principal null directions of the Weyl tensor.
In conjunction with the spin coefficient formalism \cite{Newman:1961qr},  
the algebraically special character permits a substantial  simplification in integrating Einstein's equations. 
A landmark result stemming from this property is exemplified in the work of Kinnersley, which enables us to obtain the exhaustive list of the 
Petrov-D vacuum solutions in a closed form \cite{Kinnersley:1969zza}. 

Robinson and Trautman have studied a spacetime admitting an expanding congruence of 
null geodesics which is shear-free and twist-free \cite{RT,RT2}. The Robinson-Trautman family encompasses various physically important classes of 
spacetimes, including the static black holes, the C-metric and other radiative solutions.
Since the discovery of the solution, the  Robinson-Trautman family has been proven to offer an intriguing arena for exploring 
various issues of gravitational radiation, black holes and asymptotic structure of spacetimes.

The Robinson-Trautman solution is characterized by a solution to the fourth-order nonlinear partial differential equation. 
Tod pointed out that the Robinson-Trautman equation is identified as the Calabi flow~\cite{Tod}. 
The question regarding the existence of solution  was initially elaborated within the framework of  the linear approximation
\cite{FN}. It turns out that the  perturbations decay exponentially at large retarded time, causing the spacetime to settle down to the 
Schwarzschild solution. Luk\'acs et al. exploited the Lyapunov functional to demonstrate the validity of this scenario 
in the nonlinear regime \cite{Lukacs:1983hr}. A power series solution was investigated by 
\cite{Vandyck,Vandyck2}, while the asymptotic behavior at spatial infinity was analyzed by \cite{Schmidt}. 
The prolongation structure and the B\"acklund transformation of the Robinson-Trautman equation were
explored in-depth by \cite{Glass,Hoenselaers,Hoenselaers2}. The global solution was  extensively addressed by several authors\cite{Rendall,Singleton,Chrusciel:1991vxx,Chrusciel:1992rv,Chrusciel:1992tj}. Their work shows that the 
Robinson-Trautman solution exists globally for generic, arbitrarily strong and smooth initial data, providing a conclusive proof 
that the solution eventually converges to the Schwarzschild metric. However, we must note that the existence of radiation comes at a price. 
A tantalizing physical consequence is that the 
extension through the event horizon is feasible only with a finite degree of 
differentiability.

The nonvacuum generalization has also been discussed from various points of view.  
The study of the Robinson-Trautman solution in Einstein-Maxwell system has primarily focused upon 
the aligned case, for which the repeated eigenvector of the Weyl tensor is also an eigenvector of the Maxwell field \cite{Kozameh:2006hk}. 
In the case of a positive cosmological constant, Bicak and Podolsky carried out the thorough analysis for 
the black hole formation and the cosmic no-hair conjecture \cite{Bicak:1995vc,Bicak:1997ne}. 
In the case of anti-de Sitter (AdS), the field theory interpretation has been elucidated using 
the gauge/gravity duality  \cite{BernardideFreitas:2014eoi,Bakas:2014kfa}. 
A systematic study of the Robinson-Trautman solutions in higher dimensions can be found in \cite{Podolsky:2006du,Ortaggio:2007hs,Ortaggio:2014gma}. 
All algebraically special subcases of Robinson-Trautman family have been completely classified by
\cite{Podolsky:2016sff} without specifying the matter contents.

The main aim of the current work is to implement a comprehensive study of the Robinson-Trautman family in the presence of a scalar field. 
Since the scalar fields appear in a natural fashion in supergravity and string theory,  the extensive analysis is certainly of physical significance. 
The authors of \cite{Gueven:1996zm} discussed the Robinson-Trautman solution 
in Einstein-Maxwell-dilaton gravity. It turns out that only a special coupling constant arising from string theory
admits a dynamical solution. The Robinson-Trautman solution with a massless scalar field has also been addressed by \cite{Tahamtan:2015sra,Tahamtan:2016fur} for some restricted class.  The goal of the present paper is to extend  the analysis of  \cite{Tahamtan:2015sra,Tahamtan:2016fur} into several directions. 

We successfully complete the program for classification of spacetimes admitting a congruence of expanding null geodesics without shear and rotation in Einstein gravity 
sourced by a massless scalar field. It turns out that the solutions fall into three distinct classes: (I-a) a natural extension of the Robinson-Trautman family endowed with a scalar hair, (I-b) a novel non-asymptotically flat solution, and (II) a dynamical solution possessing ${\rm SO}(3)$, ${\rm ISO}(2)$ or ${\rm SO}(1,2)$ symmetry. 
The class (I-a) solution is specified by a single function obeying the analogue of the vacuum Robinson-Trautman equation. 
We find that the solution to the corresponding equation can be generated by a solution to the Ricci flow equation. 
 In the case of the class (I-b),  the solution is characterized by two functions obeying a pair of Perelman's Ricci flow equations. 
The possibility that the appearance of Ricci flow has deeper mathematical meanings or implications is alluring.
We give an exhaustive list of all explicit solutions falling into Petrov type D for classes (I-a) and (I-b). 
Furthermore, we derive the neutral Robinson-Trautman solution to the  ${\cal N}=2$ gauged supergravity with the prepotential $F(X) =-iX^0X^1$, 
based on the class (I-a) solution in the massless case. By flipping the sign of the kinetic term of the scalar field, the present solution leads to a time-dependent wormhole with an instantaneous spacetime singularity. Although the general solution is unavailable for class (II), we find a new solution via AdS/Ricci-flat correspondence from the AdS-Roberts solution.

We organize our paper as follows. 
The next section is devoted a summary of the main features of the Robinson-Trautman solution to the 
vacuum Einstein's equations with a cosmological constant. Section \ref{RT:scalar} describes the 
Einstein-scalar system and puts constraints on the metric  admitting a congruence of null geodesics which is shear-free and 
hypersurface-orthogonal.
In section \ref{sec:classification}, we present a classification of Robinson-Trautman spacetimes for the massless scalar field. 
The asymptotically AdS Robinson-Trautman solution in supergravity is given in section~\ref{RT:AdS},  
in which a complete classification of Petrov D spacetimes is also presented. 
In section \ref{sec:D}, we explore the physical properties and causal structure of the 
Petrov-D solutions in detail. Section \ref{sec:summary} concludes our paper with some future outlooks. 
Appendix~\ref{sec:RF} comprises reviews on Ricci flow for concreteness. 
More physical discussions for class (I-b) solution are found in appendix \ref{sec:classIb}. 
A new class (II) solution is constructed in appendix \ref{sec:classII} with the aid of the AdS/Ricci-flat correspondence. 

Our conventions of curvature tensors are 
$[\nabla _\rho ,\nabla_\sigma]V^\mu ={R^\mu }_{\nu\rho\sigma}V^\nu$ 
and ${R}_{\mu \nu }={R^\rho }_{\mu \rho \nu }$.
The Lorentzian metric is taken to be the mostly plus sign, and 
Greek indices run over all spacetime indices. 
To keep simplicity of equations, we work in units $8\pi G=c=1$.

\section{$\Lambda$-vacuum Robinson-Trautman family}

Before going into the detail of the Robinson-Trautman solution in Einstein-scalar system, 
we  first revisit  the 
Robinson-Trautman family in Einstein-$\Lambda(<0)$ system in this section. 

The Robinson-Trautman family of solutions is characterized by the existence of 
a shear-free and twist-free null geodesic congruence with a nonvanishing expansion \cite{RT,RT2}. In terms of the Newman-Penrose quantities \cite{Newman:1961qr,Chandrasekhar:1985kt,Stewart:1990uf}, this condition is expressed as\footnote{
Since we are working on the mostly minus sign of metric, we follow the convention of \cite{Stephani:2003tm}} 
\begin{subequations}
\label{RTcond1}
\begin{align}
\kappa\equiv&\, -m^\mu l^\nu \nabla_\nu l_\mu =0 \,, \\
\sigma\equiv&\, -m^\mu m^\nu \nabla_\nu l_\mu =0\,, \\
\rho\equiv &\,-m^\mu\bar m^\nu \nabla_\nu l_\mu =\bar \rho \ne 0\,. 
\label{rho}
\end{align}
\end{subequations}
Here we work in the null tetrad $g_{\mu\nu}=-2l_{(\mu}n_{n)}+2m_{(\mu}\bar m_{\nu)}$ 
with $l^\mu n_\mu =-1=-m^\mu\bar m_\mu$ and $l^\mu$ corresponds to the generator of null geodesics. 
In the $\Lambda$-vacuum case, 
the Goldberg-Sachs theorem \cite{GS} immediately tells us that the spacetime is algebraically special
\begin{subequations}
\label{RTcond2}
\begin{align}
\label{}
\Psi_0\equiv &\, C_{\mu\nu\rho\sigma}l^\mu m^\nu l^\rho m^\sigma =0 \,, 
\\
\Psi_1\equiv &\, C_{\mu\nu\rho\sigma}l^\mu n^\nu l^\rho  m^\sigma=0 \,, 
\end{align}
\end{subequations}
where $C_{\mu\nu\rho\sigma}$ is the Weyl tensor. 

In Einstein-$\Lambda$ system, these geometric properties are utilized to 
obtain the most general solution to Einstein's equations. 
The Robinson-Trautman metric reads \cite{RT,RT2}
\begin{align}
\label{vacuum}
\D s^2 =-\left(K-2 r \partial_u \ln P -\frac{2m}{r}+g^2 r^2 \right)\D u^2-2 \D u \D r+\frac{2r^2}{P^2}\D z \D \bar z \,, 
\end{align}
where $m$ and $g$ are constants representing respectively the mass 
and the inverse of AdS radius $\Lambda=-3g^2$. $K=K(u, z,\bar z)$ is defined in terms of $P=P(u, z, \bar z)$ by 
\begin{align}
\label{K}
K(u, z, \bar z)\equiv 2P(u, z, \bar z)^2 \partial_z\partial_{\bar z}\ln P(u, z, \bar z)\,. 
\end{align}
It follows that $K$ represents the Gauss curvature for the (possibly $u$-dependent) two-dimensional metric 
\begin{align}
\label{g2}
g^{(2)}_{ij}\D x^i \D x^j = 2 P^{-2}\D z \D \bar z \,, \qquad x^i=(z, \bar z) \,. 
\end{align}
We will consistently use the foregoing notation (\ref{K}) and (\ref{g2}) throughout the paper. 
For the satisfaction of vacuum Einstein's equations, $P$ must obey the following Robinson-Trautman equation~\cite{RT,RT2}
\begin{align}
\label{RTeqvacuum}
\Delta_2 K+12 m \partial_u (\ln P)=0\,,
\end{align}
where $\Delta_2 =2P^2 \partial_z\partial_{\bar z}$ denotes the Laplace operator for the 
two-dimensional metric  (\ref{g2}).
The solution admits a shear-free and twist-free null geodesic congruence generated by 
$l=\partial/\partial r$.  The $r=0$ is a central curvature singularity for $m\ne 0$.

The Robinson-Trautman equation (\ref{RTeqvacuum}) is the parabolic differential equation for $P$ of fourth-order in spatial derivatives. 
It is enlightening to cast this equation into a covariant form
\begin{align}
\label{RTflow}
\frac{\partial }{\partial u}g_{ij}^{(2)}=\frac 1{12 m }\left(\Delta_2 R^{(2)}\right)g_{ij}^{(2)} \,, 
\end{align}
where  
$R^{(2)}=2K $ is the scalar curvature for the two-dimensional metric (\ref{g2}). 
Since $g_{ij}^{(2)}$ is two-dimensional, the above equation 
precisely accords with the Calabi flow~\cite{Tod},
\begin{align}
\label{CF}
\frac{\partial }{\partial t}g_{z \bar z}^{(2)}=\frac{\partial^2 R^{(2)}}{\partial z\partial \bar z } \,, 
\end{align}
where $t=u/(6m)$ is a flow parameter and 
$g_{z\bar z}^{(2)}$ is the K\"ahler metric in the basis of complex coordinates $z$.
This provides a volume preserving deformation within a given K\"ahler class of metrics. 
The stationary point of the flow is given by $\partial_u \ln P=0$, viz., $P=P(z, \bar z)$. This includes the 
topological Schwarzschild-AdS family for which the two-dimensional space $2P^{-2}\D z \D \bar z$ 
is a space of constant Gauss curvature $K(u, z, \bar z)=k$, which can be normalized as $k=0, \pm 1$.  
We can exploit the representative of the constant curvature space $2P^{-2}\D z \D \bar z$  as 
\begin{align}
\label{dSigma}
\D \Sigma_k^2 =\frac{2\D z\D \bar z}{(1+kz\bar z/2)^2}=\D \theta ^2 +\left(\frac 1{\sqrt k}\sin (\sqrt k \theta)\right)^2 \D \varphi^2 \,, 
\end{align}
where we have set $z=\sqrt{2/k}\tan(\sqrt k\theta/2)e^{i\varphi}$ at the second equality. 
Then, the metric is reduced to a familiar form 
\begin{align}
\label{}
\D s^2= -\left(k-\frac{2m}{r}+g^2 r^2\right)\D u^2 -2 \D u \D r +r^2 \D \Sigma_k^2 \,. 
\end{align}

The Robinson-Trautman solution (\ref{vacuum}) with $k=1$ and $g=0$  incorporates black hole formation with almost spherical  gravitational radiation
 in the asymptotically flat spacetime. 
For an arbitrary prescribed initial data $P_0=P(u_0, z, \bar z)$ on the surface $u=u_0$, 
it has been proven that the solution to (\ref{RTeqvacuum}) exist globally for $u\ge u_0$. 
Luk\'acs et al. exploited the Lyapunov functional to  prove  that the vacuum Robinson-Trautman solution asymptotically settles down to the Schwarzschild spacetime
\cite{Lukacs:1983hr}. 
Writing $P(u, z, \bar z)\equiv (1+z\bar z/2) \mathfrak f(u, z, \bar z)$, 
the precise fall-off rate for $u\to \infty$ is obtained by solving (\ref{RTeqvacuum})
asymptotically and reads  \cite{Chrusciel:1992tj}
\begin{align}
\mathfrak f=&\,\sum_{i,j \ge 0} \mathfrak f_{(i,j)}u^j e^{-2iu/m} 
\notag \\
=&\, 1+\mathfrak f_{(1,0)}e^{-2u/m}+\cdots +\mathfrak f_{(14,0)}e^{-28u/m}
+\mathfrak f_{(15,0)}e^{-30u/m}+\mathfrak f_{(15,1)}u e^{-30u/m}+\cdots \,,
\label{asyf}
\end{align}
where $\mathfrak f_{(i,j)}$ are real smooth functions of $z, \bar z$, satisfying 
$\mathfrak f_{(i,j)}=0$ for {$i\le 14$} and $j\ge 1$.\footnote{
The time dependence $e^{-2u/m}$ of the first nontrivial term in (\ref{asyf}) stems from the 
$l=2$ mode of linearized solution, when it is expanded by spherical harmonics $Y_{lm}$ \cite{Schmidt}. 
} 
Due to the exponential damping, the solution promptly approaches 
to the Schwarzschild spacetime as $u\to \infty$  \cite{Chrusciel:1992tj}. 
It follows that the $u=+\infty$ surface is identified as a locus of the event horizon, 
even for this dynamical setting.

It deserves to comment that the extension across the event horizon $u=+\infty$ is not analytic, 
due to the presence of radiation. 
To demonstrate this, it is of convenience to introduce Kruskal-like coordinates
\begin{align}
\label{Kru}
\hat u=- e^{-u/(4m)} \,, \qquad \hat v=e^{v/(4m)}\,, 
\end{align}
where $v=u+2r+4m \ln [(r-2m)/(2m)]$. In terms of $\hat u$, 
the asymptotic solution (\ref{asyf}) is rephrased as 
\cite{Chrusciel:1992rv}
\begin{align}
\label{}
\mathfrak f=1+\mathfrak f_{(1,0)}\hat u^8+\cdots +\mathfrak f_{(14,0)}\hat u^{112}
+\mathfrak f_{(15,0)}\hat u^{120}+\mathfrak f_{(15,1)}(-4m\ln |\hat u|) \hat u^{120}+\cdots \,,
\end{align}
It follows that 
the presence of the term $\mathfrak f_{(15,0)}$ with logarithm prevents the solution to be smooth at 
$\hat u=0$.  Since $\mathfrak f$ is at most $C^{119}$, the full solution is $C^{117}$. 
For $\Lambda<0$,  the term $1/(4m)$ in (\ref{Kru}) should be replaced by the surface gravity for the 
Schwarzschild-AdS spacetime, leading to  worse degree of differentiability.
For instance, 
 the metric is not even first-order differentiable at $u=+\infty$ for large AdS black holes with $m^2 g^2>4/27$
 \cite{Bicak:1997ne}.

\section{Robinson-Trautman solutions in Einstein-scalar system}
\label{RT:scalar}

\subsection{Einstein-scalar system with a potential}

Let us consider the four-dimensional Einstein's gravity with a real scalar field $\phi$ described by the Lagrangian
\begin{align}
\label{Lag}
\ma L=R -\varepsilon (\nabla\phi)^2 -2  V(\phi)\,,
\end{align}
where 
$R$ is the scalar curvature and $V(\phi)$ is the scalar potential. The scalar field is normal for 
$\varepsilon=+1$ and phantom for $\varepsilon=-1$. We keep $\varepsilon$ throughout the paper 
to emphasize the consequence of normal/phantom aspect. The field equations derived from the above Lagrangian read
\begin{align}
\label{Eineq}
E_{\mu\nu}\equiv R_{\mu\nu}-\frac 12 R g_{\mu\nu}-T_{\mu\nu}=0 \,, \qquad 
E^{(\phi)}\equiv \varepsilon \nabla^2 \phi-\partial_\phi V=0 \,,
\end{align}
where $T_{\mu\nu}$ is the stress energy tensor of the scalar field
\begin{align}
\label{}
T_{\mu\nu}=\varepsilon \left(\nabla_\mu \phi \nabla_\nu \phi -\frac 12 (\nabla \phi)^2 g_{\mu\nu}\right)-V(\phi)g_{\mu\nu} \,. 
\end{align}

In this paper, 
we focus on a theory in which the potential $V(\phi)$ of the scalar field is expressed in terms of a real-valued subsidiary function 
\begin{align}
\label{W}
W(\phi)=\frac{g}{2} \cosh \left(\sqrt{\frac \varepsilon 2}\phi \right)\,.
\end{align}
as 
\begin{align}
\label{pot:superpotential}
V(\phi)=4\left[2\varepsilon \left(\frac{\partial}{\partial \phi}W(\phi)\right)^2-3 W(\phi)^2 \right]
=-g^2 \left[2+\cosh (\sqrt{2\varepsilon}\phi)\right]\,. 
\end{align}

For $\varepsilon=+1$, the present model is obtained by the truncation of the ${\cal N}=2$ supergravity with the prepotential $F(X)=-i X^0X^1$
(see, e.g., \cite{Klemm:2011xw,Colleoni:2012jq,Gnecchi:2013mja}), 
\begin{align}
\label{FIV}
V(\phi)=-{2} \left(g_0^2 e^{\sqrt 2\phi}+4g_0g_1 +g_1^2e^{-\sqrt 2\phi} \right)\,,
\end{align}
where $g_0$ and $g_1$ are Fayet-Iliopoulos gauge coupling constants. 
By choosing $g_0=g_1=g/{2}$, we recover (\ref{pot:superpotential}). 
In this context, $W=(g/2)\cosh (\phi/\sqrt 2)$ represents the superpotential and $V(\phi)$ admits a unique 
AdS vacuum at the origin $\phi=0$.

For $\varepsilon=-1$, the model  is formally obtained by analytic continuation $\phi\to i\phi$ 
of the $\varepsilon=+1$ case. In spite of this simple prescription, one can observe a qualitative change of the 
potential $V(\phi)$  as  it now admits an infinite number of extrema \cite{NT2}
\begin{align}
\label{cpts}
\phi_n= \sqrt{2}n \pi \,, \qquad 
\ti \phi_n= \sqrt{2}\left(n+\frac 12 \right)\pi \,, \qquad 
n\in \mathbb Z \,.
\end{align}
The critical points $\phi_n$ correspond to the local minima of the potential satisfying 
\begin{align}
\label{AdSradii}
V(\phi_n)=-3 g^2 \,, \qquad 
m^2_n\equiv \varepsilon V''(\phi_n)=-2g^2 \,. 
\end{align}
Thus, $g$ corresponds to the inverse of the AdS radius at each extremum. 
These critical points $\phi_n$ also extremize the superpotential $W'(\phi_n)=0$, while at the 
local maxima we obtain $W'(\ti \phi_n)\ne 0$ and 
\begin{align}
\label{}
V(\ti \phi_n)=-3 \ti g^2 \,, \qquad \ti g \equiv  \frac{g}{\sqrt 3}\,, \qquad 
\ti m^2_n\equiv \varepsilon V''(\ti \phi_n)=6 \ti g^2 \,. 
\end{align}

Our new class (I-a) solution with the potential, as will be derived in the next section, asymptotically tends to 
AdS at the origin $\phi=0$. Since the mass spectrum (\ref{AdSradii}) lies in the characteristic range
$-(9/4)g^2\le m^2 < -(5/4)g^2$, the slower fall-off mode of the scalar field is also normalizable~\cite{Ishibashi:2004wx}. 
Specifically, the scalar field obeys the mixed Robin boundary conditions. 
This entails the slower fall-off for the metric around AdS vacuum, as opposed the Dirichlet boundary conditions~\cite{Hertog:2004dr,Henneaux:2006hk}.

\subsection{Metric form}

We shall now consider the Robinson-Trautman solutions in the Einstein-scalar system, i.e., 
solutions for field equations (\ref{Eineq}) admitting a shear-free and twist-free null geodesic congruence satisfying (\ref{RTcond1}). 
Since the Goldberg-Sachs theorem \cite{GS} is inapplicable for the present system, 
the algebraically special feature does not automatically follow from (\ref{RTcond1}). 
We are now going to describe the conditions under which the natural generalization of the Robinson-Trautman class of solutions 
is achieved.

First of all, the Newman-Penrose equation [(7.21b) of \cite{Stephani:2003tm}] implies $\Psi_0=0$ under $\sigma=\kappa=0$.
Observe that conditions in (\ref{RTcond1}) and $\Psi_0=0$ are all preserved under the change of null tetrad basis of class-I
$\{l^\mu\to l^\mu, m^\mu\to m^\mu+a_{\rm I}l, n^\mu\to n^\mu+\bar a_{\rm I}m^\mu+a_{\rm I}\bar m^\mu+|a_{\rm I}|^2 l^\mu\}$, 
and class III $\{l^\mu\to a_{\rm III} l^\mu$, 
$n^\mu\to a_{\rm III}^{-1} n^\mu$, 
$m^\mu\to e^{i\vartheta_{\rm III}}m^\mu\}$, 
where $a_{\rm I}$ is complex while $a_{\rm III}$ and $\vartheta_{\rm III}$ are real. 
Using class III freedom, one can enforce a gauge 
\begin{align}
\label{}
\epsilon \equiv -\frac 12 (n^\mu l^\nu \nabla_\nu l_\mu -\bar m^\mu l^\nu \nabla_\nu m_\mu)=0\,. 
\end{align}
In light of this,  the null geodesics are affinely parameterized $l^\nu\nabla_\nu l^\mu=0$.

Under the class I null rotation, $\tau\equiv  -m^\mu n^\nu \nabla_\nu l_\mu$ varies  as $\tau \to \tau+a_{\rm I} \rho$. 
Since $\rho $ is nonvanishing, it is always possible to impose the following gauge condition for class I change of null tetrad basis
\begin{align}
\label{RTcondgauge}
\tau= 0\,. 
\end{align}
This gauge has been also employed to derive the canonical form of the vacuum Robinson-Trautman solution [(27.6) of \cite{Stephani:2003tm}].

In the presence of the scalar field, the Weyl tensor is not algebraically special, since 
the geometric optical conditions (\ref{RTcond1}) do not directly impose a constraint
on the spacetime curvature. Although the pursuit of algebraically general solutions is worth investigating, 
we enforce in this paper the algebraically special condition $\Psi_1=0$. 
Furthermore, we require the following two conditions
\begin{subequations}
\label{RTcond3}
\begin{align} 
\Phi_{01}\equiv &\,\frac 12 R_{\mu\nu}l^\mu m^\nu =0\,,
\\
\Phi_{02}\equiv & \, \frac 12 R_{\mu\nu}m^\mu m^\nu =0\,.
\end{align}
\end{subequations}
These conditions put constraints on the possible form of the scalar field configuration, 
via Einstein's equations. As we shall see shortly, this requires the scalar field to be 
``angular independent.''  
Note also that $\Phi_{00}\equiv \frac 12  R_{\mu\nu}l^\mu l^\nu$ is nonvanishing, 
since the scalar field is not aligned generically.


One can derive further constraints on spin coefficients from the above conditions.
Combined with the Newman-Penrose equation [(7.21c) of \cite{Stephani:2003tm}], 
we obtain
\begin{align}
\label{addcond1}
\pi\equiv \bar m^\mu l^\nu \nabla_\nu n_\mu =0 \,. 
\end{align}
This condition implies that  other null tetrads are parallelly  propagated along  $l^\mu$ as 
\begin{align}
\label{ppframe}
l^\nu \nabla_\nu l^\mu=0\,, \qquad 
l^\nu \nabla_\nu n^\mu=0 \,, \qquad l^\nu \nabla_\nu m^\mu=0 \,.
\end{align}
Due to $\Phi_{00}\ne 0$, 
the Bianchi identity for the Riemann tensor [(7.32b) of \cite{Stephani:2003tm}] entails
\begin{align}
\label{addcond2}
\lambda\equiv \bar m^\mu \bar m^\nu \nabla_\nu n_\mu =0\,. 
\end{align}
Accordingly, the dual null vector $n^\mu$, which is not geodesic, is also shear-free. 
Turning the logic around, one can deduce $\Phi_{01}=0$ from the assumption $\pi=0$
and similarly  $\Phi_{02}=0$  from $\lambda=0$, provided $\Psi_1=0$.


Let us now adapt the privileged local coordinates.  
Letting $r$ denote the affine parameter of the null geodesics, we 
have $l^\mu=(\partial/\partial r)^\mu$. 
One finds from $\kappa=\epsilon=\rho-\bar\rho=0$ that $l_\mu$ is hypersurface-orthogonal $l_{[\mu}\nabla_{\nu}l_{\rho]}=0$. 
This indicates that there exist $r$-independent functions 
$\mas H$ and $u$ such that  $l_\mu=-\mas H\nabla_\mu u$. 
Performing further class III change of null tetrad basis composed of $r$-independent $a_{\rm III}$, 
one can set $\mas H=1$  without loss of generality.

One can also ascertain that 
the complex vector $m_\mu$ becomes hypersurface-orthogonal $m_{[\mu}\nabla_{\nu}m_{\rho]}=0$, implying that
 there exist complex scalars  $\mas P$ and $z$ such that 
$\bar m=\mas P^{-1}\D z$, where $z$ is independent of $r$.  
Because of the gauge $\epsilon=0$, it follows that  $\mas P$ is real. 
Lastly, the null vector $n^\mu$ can be parametrized as
$n_\mu=-(\nabla_\mu r+\frac 12 H \nabla _\mu u+\varpi \D z+\bar \varpi \D \bar z)$, 
where $H$ is real and $\varpi$ is complex. 

The 
four functions ($r$, $u$, $z$, $\bar z$) span the local coordinates of the spacetime, where $u$ can be regarded as a retarded time. 
Since $\varpi$ is independent of $r$ as a result of $\tau=\pi=0$,  
one can resort to the transformation $r\to r+w_0(u, z,\bar z)$ to 
set $\varpi=0$. Inspecting $\Phi_{01}=0$, one obtains $\partial_z\partial_r(\ln \mas P)=0$,
which is integrated to give $\mas P=P(u, z, \bar z)/S(u, r)$. 
A part of Einstein's equations $E_{\mu\nu}m^\mu m^\nu=0$ 
gives rise to $\phi=\phi(u,r)$. 
The solution is therefore put into the form
\begin{align}
\label{RTmetric}
\D s^2=-2 \D u \D r -H(u,r, z, \bar z) \D u^2+2 \frac{S(u,r) ^2}{P(u,z, \bar z)^2}\D z \D \bar z \,, \qquad 
\phi=\phi(u, r) \,,
\end{align}
where $H$, $S$ and $P$ are all real functions.  The convergence $\rho$  of the null geodesics given by (\ref{rho}) is 
\begin{align}
\label{rho2}
\rho =-\frac{\partial _ r S}{S} \,. 
\end{align}
Here, $r$ is the affine parameter, $u$ is the retarded time and ($z, \bar z$) span the deformed maximally symmetric space. 
In the ensuing sections, we shall determine these functions by imposing 
Einstein's field equations (\ref{Eineq}).

\section{Classification of solutions without the scalar potential}
\label{sec:classification}

To begin with, we give a complete classification of the 
Robinson-Trautman class of solutions without the scalar potential $g=0$, i.e., for the massless scalar field. 
The solution can be written in local coordinates ($u, r, z, \bar z$) as
(\ref{RTmetric}). In the forthcoming discussion, we impose Einstein's equations (\ref{Eineq}) 
to acquire the explicit form of solutions.
Both of the ordinary scalar $\varepsilon=+1$ and the phantom scalar $\varepsilon=-1$
will be analyzed.

From $E_{uz}=E_{u\bar z}=0$, we immediately obtain  
\begin{align}
\label{f0}
H(u,r, z, \bar z)=H_0(u,r)+H_1(u, z, \bar z)-2r\partial_u \ln P(u, z, \bar z)\,.
\end{align}
The equation coming from $\partial_z[S^2(2E_{ur}-H E_{rr})]=0$ gives rise to 
\begin{align}
\label{Azeq}
\partial_{\bar z}K +2 \partial_r \left(r S \partial_r S-S^2\right) \partial_{\bar z}\partial_{u}\ln P
-\partial_r \left(S\partial_r S\right)\partial_{\bar z}H_1=0 \,,
\end{align}
where $K\equiv 2P^2 \partial_z\partial_{\bar z}\ln P$ as before (\ref{K}).
Since the first term is insensitive to $r$, the  ensuing discussion divides into two cases, according to\footnote{
Alternatively, one can classify the spacetime by the following fashion. Equation (\ref{Azeq}) and its complex conjugation 
can be cast into the linear system
\begin{align}
\label{}
\mas A(u,z,\bar z) \left(
\begin{array}{c}
X_1(u,r)    \\
X_2(u,r)    
\end{array}
\right)=-\left(
\begin{array}{c}
\partial_z K(u,z, \bar z)    \\
\partial_{\bar z} K(u,z, \bar z) 
\end{array}
\right)\,,\notag \qquad 
\mas A(u,z,\bar z) \equiv \left(
\begin{array}{cc}
\partial_{z}p(u,z,\bar z)     &  \partial_{z}H_1 (u,z,\bar z)  \\
\partial_{\bar z}p  (u,z,\bar z)   &  \partial_{\bar z}H_1 (u,z,\bar z)
\end{array}
\right)\,,
\end{align}
where $X_1(u,r)=2\partial_r(rS\partial_r S-S^2)$, $X_2(u,r)=-\partial_r (S \partial_r S)$ and $p\equiv \partial_u \ln P(u,z,\bar z)$. 
The space of solutions for this system divides into (I') ${\rm det}\mas A\ne 0$ and  (II') ${\rm det}\mas A\ne 0$. 
For the case (I'), one can invert $\mas A$ to deduce $X_i=X_i(u)$, which is included in--but not identical to--class (I), 
because of an extra condition ${\rm det}\mas A\ne 0$.  
For the case (II'), one can check that the all possible cases end up with class (I) or class (II). Here, 
we refrain from classifying spacetimes according to (I') and (II'), since classes (I) and (II) are intricately intertwined, making it difficult to distinguish between them.
}
\begin{align}
\notag
\textrm{Class I}:&~~\textrm{$\partial_{\bar z}\partial_{u}\ln P$ and $\partial_{\bar z}H_1$ do not vanish simultaneously}. \\
\textrm{Class II}:&~~\partial_{\bar z}\partial_{u}\ln P=\partial_{\bar z}H_1= 0 \,. \notag
\end{align}
In the case of Class I, the coefficients of $\partial_{\bar z}\partial_{u}\ln P$ or 
$\partial_{\bar z}H_1$ in (\ref{Azeq}) must be a function only of $u$. In any case, 
the function $S(u,r)$ in Class I takes the form
\begin{align}
\label{S0}
S=\sqrt{U_1(u)r^2+U_2(u)r-\varepsilon U_0(u)} \,.
\end{align}
Here we have implanted $\varepsilon$ in front of $U_0(u)$ for the sake of later convenience. 
The analysis of Class I further splits into two cases, according to 
(I-a) $U_1(u)\ne 0$ or (I-b)  $U_1(u)= 0$ and $U_2(u)\ne 0$. 
We shall not consider the $U_1(u)=U_2(u)=0$  case, since this falls outside the Robinson-Trautman class
$\rho \ne 0$ [see (\ref{rho2})].

\subsection{Class I-a}

For $U_1(u)\ne 0$, one can exploit the freedom $r\to r+w_0(u)$ to 
achieve $U_2(u)=0$. 
To keep the Lorentzian signature in the sufficiently large $r$ region, $U_1>0$ is assumed 
in the hereafter. Along with (\ref{S0}), 
integration of $ 2E_{ur}-H E_{rr}=0$ gives  
\begin{subequations}
\begin{align}
\label{H0}
H_0(u,r)=&\, \frac{h_0(u)}{r}+\frac{r U_1'(u)}{U_1(u)}\,, \\
\label{H1}
H_1(u, z, \bar z)=&\, \frac{K(u, z, \bar z)}{U_1(u)} \,, 
\end{align}
\end{subequations}
where the prime denotes the differentiation with respect to $u$, and $K(u, z, \bar z)$ stands for the 
shorthand notation of $2P^2\partial_z\partial_{\bar z}\ln P$ as in the vacuum case (\ref{K}).

For $U_0=0$, it turns out that the only allowed solution is $\phi={\rm const}$, 
which we exclude from our discussion. For $U_0\ne 0$, 
$E_{rr}=E_{z\bar z}=0$ are combined to give
\begin{subequations}
\label{phiur}
\begin{align}
\partial_r \phi=&\,\pm \frac{\sqrt{2 U_1U_0}}{U_1r^2-\varepsilon U_0} \,, \\
\partial_u \phi=&\, \pm \frac{1}{\sqrt{2U_1U_0}}\frac{r^4(U_0U_1'-U_0'U_1)-h_0U_0(U_1r^2-\varepsilon U_0)}{r^3(U_1r^2-\varepsilon U_0)}\,. 
\end{align}
\end{subequations}
It follows that $U_0>0$ is demanded for the reality of $\phi$. 
Compatibility $(\partial_r\partial_u-\partial_u\partial_r)\phi=0$ of these first order equations  gives $h_0(u)=0$. 
Now, the scalar field equation $\nabla^2 \phi=0$ is satisfied and $E_{uu}=0$ yields the following master equation
\begin{align}
\label{Keq0}
\Delta_2 K+2\varepsilon P^2 \sqrt{U_0 U_1} \partial_u \left[\sqrt{ \frac{U_1}{U_0}}\partial_u \left(\frac{U_0}{P^2}\right)\right]=0\,,
\end{align}
where $\Delta_2 =2P^2 \partial_z \partial_{\bar z}$ is the Laplace operator for the metric $2P^{-2}\D z \D \bar z$. 
This is a nonlinear partial differential equation of fourth-order in spatial derivatives for $P$, corresponding to the 
Robinson-Trautman equation (\ref{RTeqvacuum}) in the vacuum case.  It follows that 
the metric is now reduced to 
\begin{align}
\label{sol0}
\D s^2=-2\D u\D r - \left(\frac{K}{U_1}+\frac{rU_1'}{U_1}-2r\partial_u \ln P\right)\D u^2
+\frac{2(U_1r^2-\varepsilon U_0)}{P^2}\D z \D \bar z \,. 
\end{align}

Let us finally determine the profile of the scalar field. 
By choosing the sign in (\ref{phiur}) and the integration constant appropriately, 
we obtain
\begin{align}
\label{phip}
\phi=\frac 1{\sqrt 2} \ln \left(\frac{\sqrt{U_1(u)}r-\sqrt{U_0(u)}}{\sqrt{U_1(u)}r+\sqrt{U_0(u)}}\right)\,.
\end{align}
for the ordinary scalar field $\varepsilon=+1$ and 
\begin{align}
\label{phians}
\phi=\sqrt 2 \left[\frac \pi 2-\arctan \left(\sqrt{\frac{U_1(u)}{U_0(u)}}r\right)\right]\,,
\end{align}
for the phantom scalar field $\varepsilon=-1$.
In either case, the $U_0\to 0+$ limit gives rise to the $m=g=0$ vacuum solution (\ref{vacuum}).

It should be emphasized that the solution (\ref{sol0}) is form-invariant under the reparametrization 
$(u, r, z)\to (\ti u, \ti r, \ti z)$ where 
\begin{align}
\label{w1w2}
\ti r=w_1(u)w_2(u) r \,, \qquad 
\D \ti u=\frac{\D u}{w_1(u)w_2(u)} \,, \qquad 
\ti z=\ti z(z) \,, 
\end{align}
with 
\begin{align}
\label{}
\ti U_1(\ti u)=\frac{U_1(u)}{w_1(u)^2} \,, \qquad 
\ti U_0 (\ti u)=w_2(u)^2 U_0(u) \,, \qquad 
\ti P(\ti u, \ti z, \bar{\ti z})=w_2(u)\left|\frac{\partial \ti z}{\partial z}\right|P(u, z, \bar z) \,. 
\end{align}
By exploiting this freedom, we can always work in a gauge
$U_1(u)=1$ and $U_0(u)=r_0^2$ ($r_0>0$), for which the scalar field is $u$-independent. 
However,  we have left this gauge unfixed to facilitate the classification of Petrov D solutions
in section \ref{sec:PetrovD}, and 
for easy comparison with other solutions in the literature. 

The authors of  \cite{Tahamtan:2015sra} discussed 
a restricted class of Robinson-Trautman solutions with a massless scalar satisfying $\partial_u P=0$. 
In order to recover the results in  \cite{Tahamtan:2015sra,Tahamtan:2016fur}, it is necessary to work with  the gauge $U_0(u)=r_0^2 /U_1(u)$ and then to enforce the condition $\partial_u P=0$. 
In this case, 
equation (\ref{Keq0}) gives rise to  decoupled equations
\begin{align}
\label{}
\Delta_2 K (z,\bar z)=4 \varepsilon r_0^2 \omega_1 \,, \qquad 
U_1(u)=e^{\omega_1 u^2+\omega_2 u}\,,
\end{align}
where $\omega_1$ and $\omega_2$ are constants. 
For $\omega_1\ne 0$, $\omega_2$ can be set to zero by $u\to u+{\rm const}$. 
This reproduces the result in \cite{Tahamtan:2015sra,Tahamtan:2016fur}. 

In section \ref{sec:D}, we are going to investigate the physical properties of the 
solution by imposing the gauge $U_1=1$ and $U_0=r_0^2$. With this gauge fixing, 
our ($u_{\rm here}, r_{\rm here}$) and the one ($u_{\rm there}, r_{\rm there}$) in \cite{Tahamtan:2015sra,Tahamtan:2016fur}
 are transformed by 
\begin{align}
\label{}
r_{\rm here}=e^{(\omega_1 u_{\rm there}^2+\omega_2 u_{\rm there})/2}r_{\rm there}\,, \qquad 
u_{\rm here}=\int \frac{\D u_{\rm there}}{e^{(\omega_1 u_{\rm there}^2+\omega_2 u_{\rm there})/2}} \,.
\end{align}

\subsubsection{Ricci flow}
\label{sec:IaRF}

A notable and significant departure from the $\Lambda$-vacuum case  is that 
the governing equation (\ref{Keq0}) of $P$ incorporates the second $u$-derivative and fourth spatial derivative.
The advent of the second time derivative is due primarily to the effect of the scalar field. 
To gain further insight into (\ref{Keq0}), we first notice that 
equation  (\ref{Keq0})  can be cast, in a gauge $U_1(u)=1$ and $U_0(u)=r_0^2$,  into a more suggestive form\footnote{
One can dispense with this gauge fixing by considering the  
conformally transformed metric $\ti g_{ij}^{(2)}\D x^i \D x^j=2U_0(u)P^{-2}\D z \D \bar z$
and a new flow parameter $\ti t$ by $\D \ti t/\D u=\sqrt{U_0/U_1}/2$. }
\begin{align}
\label{Keq0cov}
\frac{\partial^2}{\partial t^2} g_{ij}^{(2)} 
=-\varepsilon \big(\Delta_2 R^{(2)}\big) g^{(2)}_{ij} \,,
\end{align}
where $g_{ij}^{(2)}$ is the two-dimensional metric (\ref{g2}) as before, and 
the flow parameter $t$  is defined by $t=u/(2r_0)$. 
We are now going to illustrate that the second-order $t$-derivative equation (\ref{Keq0cov}) for $\varepsilon=+1$
can be generated by a first-order equation, whose solutions automatically meet 
the second-order equation. Quite surprisingly, the pertinent first-order system turns out   
 the Ricci flow equation \cite{Hamilton}\footnote{
 Since the present system is two dimensions, the Ricci flow is equivalent of the Yamabe flow~\cite{Yamabe}. 
 }
\begin{align}
\label{RicciFlow}
\frac{\partial}{\partial t}  g_{ij}^{(2)} =-2 R^{(2)}_{ij}= - R^{(2)}  g_{ij}^{(2)} \,. 
\end{align}
In appendix~\ref{sec:RF}, we have summarized basic properties of Ricci flows. 
In fact, another operation of the $t$-derivative to (\ref{RicciFlow}) gives
\begin{align}
\label{RicciFlowderv}
 \frac{\partial^2}{\partial t^2} g_{ij}^{(2)}
 =-  \frac{\partial  R^{(2)}}{\partial t}  g^{(2)}_{ij}
 +\big( R^{(2)}\big)^2  g_{ij}^{(2)} 
 =- \big( \Delta_2 R^{(2)}\big) g_{ij}^{(2)}\,,
\end{align}
where we have used (\ref{dotR}) at the second inequality. 
It turns out that this equation reproduces the governing equation (\ref{Keq0cov}) for $\varepsilon=+1$, i.e., 
for a normal scalar field. 
It follows that any solutions to the Ricci flow equation (\ref{RicciFlow}) give rise to the solutions to  (\ref{Keq0}). 
In other words,  the ``square root'' of  equation (\ref{Keq0}) is the Ricci flow equation 
for $\varepsilon=1$. This invokes the passage from the Klein-Gordon equation to the Dirac equation.
It is remarkable that the physical solution to Einstein's equation is related to the purely geometric flow equation. 
Some explicit solutions to the Ricci flow equation are presented in appendix~\ref{sec:RF}.

\subsection{Class I-b}

Let us next focus on the case $U_1(u)=0$ with $U_2(u)\ne 0$.  The degrees of freedom $r\to w_1(u)r+w_0(u)$ enables us 
to set $U_2(u)=\ell$ and $U_0(u)=0$, where $\ell~(>0)$ is a constant with the dimension of length. $E_{ur}=E_{rr}=0$ gives
\begin{align}
\label{}
H_0(u, r)=&\, r \phi_0'(u) \,, \\\
\label{KIb}
K(u, z, \bar z)=&\,\frac {\ell}{2}\left[\phi_0'(u)+2 \partial_u \ln P(u, z, \bar z)\right]\,.
\end{align}
Recalling $K(u, z, \bar z)\equiv 2P^2 \partial_z\partial_{\bar z}\ln P$,  equation (\ref{KIb}) is the 
partial differential equation for $P$. 
$E_{rr}=E_{z\bar z}=0$ yields 
\begin{align}
\label{}
\phi=\sqrt{\frac \varepsilon{2}} \left[\phi_0(u) +\ln \left(\frac{r}{\ell}\right)\right]\,, 
\end{align}
allowing only the ordinary scalar field $\varepsilon=+1$. 
Lastly, $E_{uu}=0$ gives rise to the governing equation for $H_1=H_1(u, z,\bar z)$ as 
\begin{align}
\label{H1eq}
\big\{\Delta_2+\ell\left[\phi_0'(u)-2\partial_u \ln P(u, z, \bar z)\right]\big\}H_1=-\ell\partial_u H_1 \,,
\end{align}
where $\Delta_2=2P^2\partial_z\partial_{\bar z}$. 
This equation is  viewed as the (time-reversed) heat diffusion equation with the thermal diffusivity $1/\ell$. 
The metric reads 
\begin{align}
\D s^2=-2 \D u \D r - \left(r \phi_0'(u)-2r\partial_u \ln P+H_1\right)\D u^2 +\frac{2\ell r }{P^2 }\D z \D \bar z \,. 
\end{align}
By the transformation 
\begin{align}
\label{}
\ti r=U(u) r\,, \qquad 
\D \ti u=\frac{\D u}{U(u)}\,, \qquad 
\ti P(\ti u, z, \bar z)=U^{1/2}(u) P(u,z, \bar z)\,, \qquad 
\ti H_1(\ti u, z, \bar{z}) =U^2(u) H_1( u, z, \bar z)\,,
\end{align}
the metric and the scalar field remain invariant, up to $\phi_0\to \phi_0+\ln U$. 
Using this freedom, 
we can work in the gauge $\phi_0=0$. 

To summarize, the class (I-b) solution is given by 
\begin{align}
\label{Ibsol}
\D s^2=-2  \D u \D r- \left(H_1-2r \partial_u \ln P \right)\D u^2 +\frac{2\ell r }{P^2 }\D z \D \bar z\,, 
\qquad \phi=\frac 1{\sqrt 2} \ln \left(\frac{r}{\ell}\right)\,,
\end{align}
where  $P=P(u, z, \bar z)$ and $H_1=H_1(u, z, \bar z)$ obey 
\begin{subequations}
\label{eqIb}
\begin{align}
\Delta_2 \ln P =&\, \ell \partial_u \ln P \,,\\
\Big(\Delta_2-2 \ell \partial_u \ln P\Big)H_1=&\,-\ell\partial_u H_1\,.
\end{align}
\end{subequations}
The remaining coordinate freedom which leaves 
the solution (\ref{Ibsol}) invariant is 
\begin{align}
\label{Ibtr}
\ti z=\ti z(z)\,, \qquad 
\ti P(u, \ti z, \bar{\ti z})=\left|\frac{\partial \ti z}{\partial z}\right|P(u, z, \bar z)\,, \qquad 
\ti H_1(u, \ti z, \bar{\ti z}) =H_1( u, z, \bar z)\,,
\end{align}

It deserves to emphasize that the class (I-b) solution (\ref{Ibsol}) is not asymptotically flat. 
Specifically, the solution fails to admit the limit in which the scalar field is trivial $\phi=0$. 
This solution is therefore intrinsic to the present system and has no counterpart in the $\Lambda$-vacuum case. 

We discuss the classification of Petrov-D solutions for class (I-b)  in appendix \ref{sec:classIb}.
On account of the offbeat asymptotic structure, the physical significance of this solution is currently unclear.

\subsubsection{Pair of Ricci flow equations}

The class (I-b) solution (\ref{Ibsol}) is specified by two functions $P$ and  $H_1$ obeying (\ref{eqIb}).  
Since this pair of equations is of new type, let us make a brief but crucial comment. 
Letting $g^{(2)}_{ij}$ denote the metric for the two-dimensional line element 
$2 P^{-2}\D z \D \bar z$ as (\ref{g2}), equations in (\ref{eqIb}) precisely accord with 
a pair of  Ricci flow equations introduced by Perelman~\cite{Perelman:2006un}
\begin{subequations}
\label{RF}
\begin{align}
\label{}
\frac{\partial}{\partial t}g^{(2)}_{ij}=&\, -2 R^{(2)}_{ij} \,, \\
\frac{\partial}{\partial t} f =&\, -\Delta _2 f +(\nabla f)^2 -R^{(2)} \,, 
\end{align}
\end{subequations}
where we have defined $f \equiv -\ln |H_1|$ and $t\equiv u/\ell$. $\nabla_i$ and 
$R^{(2)}_{ij}=\frac 12 R^{(2)}g^{(2)}_{ij}=K g^{(2)}_{ij}$ are the derivative operator and 
 the Ricci tensor for the metric $g^{(2)}_{ij}$, respectively. 
 Perelman has contrived this pair (\ref{RF}) of equations 
for the resolution of Thurston's geometrization conjecture. 
To our current understanding, this is the first example that provides a pair of  Ricci flow equations
in exact gravitational solutions.

The flow equations reveal several geometric features under time evolution. For instance, the 
topology of the two-dimensional compact surface with the metric $g_{ij}^{(2)}$ remains unaltered, as 
illustrated in (\ref{chidot}). Furthermore, the pair (\ref{RF}) of flow equations enables us 
to define some ``entropy functionals'' (\ref{calF}) and (\ref{calN}). 
These quantities may have profound physical implications for the present spacetime dynamics.


\subsection{Class II}
\label{sec:classIIder}

In the case of Class (II) $\partial_{\bar z}\partial_{u}\ln P=\partial_{\bar z}H_1= 0$, the functional form of $S(u,r)$ is 
unrestricted by (\ref{Azeq}). We see that $H_1$ is a function of $u$, which can be set to zero by absorbing into $H_0(u,r)$. 
$P$ takes the product form $P(u, z, \bar z)=P_0(u)P_1(z, \bar z)$, for which we can set $P_0=1$ by absorbing it into $S(u,r)$. 
Since $K=2P^2 \partial_z\partial_{\bar z}\ln P=2 P_1^2  \partial_z\partial_{\bar z}\ln P_1$ is a constant, 
$\D s_2^2 =2P_1^{-2}\D z \D \bar z$ must be a space of constant Gauss curvature $k=0, \pm 1$.
Thus, the solution is expressed as
\begin{align}
\label{classII}
\D s^2 =-H_0(u,r)\D u^2-2\D u\D r+S(u,r)^2\D \Sigma_k^2 \,, \qquad 
\phi=\phi(u,r) \,,
\end{align}
where $\D \Sigma_k^2$ is given by (\ref{dSigma}). The class (II) solution automatically belongs to 
Petrov-type D. 
 
$E^r{}_r+(H_0/2)E^u{}_r=0$ can be integrated once, giving 
\begin{align}
\label{}
H_0(u,r)S \partial_r S -2S \partial_u S-k r=h(u) \,. 
\end{align}
The rest of field equations are 
\begin{subequations}
\begin{align}
\varepsilon (\partial_r \phi)^2 +\frac{2\partial_r^2 S}{S}=&\, 0 \,, \\ 
\frac{k}{S^2}-\left(kr+h\right)\frac{\partial_rS}{S^3}+\frac 12 \left(
\partial_r^2 H_0+\varepsilon H_0 (\partial_r \phi)^2-2\varepsilon \partial_r \phi\partial_u \phi \right)=&\, 0 \,, \\
\varepsilon \left(\partial_u \phi-\frac 12 H_0 \partial_r \phi\right)^2-\frac 1{2S^4}\left(kr+h +\frac 12 S^2 \partial_r H_0\right)^2
+\frac{4kH_0-8h'+S^2 (\partial_r H_0)^2}{8S^2}=&\, 0\,, \\
\partial_u \left(S^2\partial_r \phi\right)-\partial_r \left(S^2(H_0 \partial_r \phi-\partial_u \phi )\right)=&\, 0 \,.
\end{align}
\end{subequations}
Unfortunately, the exact solution to these equations cannot be obtained in a closed form, unless additional conditions are imposed. 
We have already encountered this obstacle in Einstein-Maxwell-dilaton system \cite{Gueven:1996zm}. 
An exceptional case is that the solution is self-similar, for which the system reduces to 
a set of ordinary differential equations~\cite{Roberts:1989sk}. 
Despite this obstacle, 
we present a new kind of Class (II) solution with the help of AdS/Ricci-flat correspondence. 
We defer the details to Appendix \ref{sec:classII}.

\section{Solutions with the scalar potential}
\label{RT:AdS}

\subsection{Robinson-Trautman family in gauged supergravity}
\label{sec:ansatz}

For the $g\ne 0$ case with the nonvanishing scalar potential, 
the procedure for solving Einstein's equations in a sequential order
does not work, as we have demonstrated for the massless case in the previous section. 
Moreover, the solution generating method is defective by the presence of the scalar potential
\cite{Klemm:2015uba}. Thus, exact solutions with $g\ne 0$ are hardly ever obtainable, 
despite widespread motivations and applications in holography. 
We therefore adapt the sophisticated ansatz-based approach to construct $g\ne 0$ solutions. 

We suppose that the 
metric (\ref{RTmetric}) with (\ref{f0}), (\ref{S0}), (\ref{H1}), (\ref{Keq0}), (\ref{phip}) and (\ref{phians}) remain invariant 
for $g\ne 0$. Under this restriction,  
we require  Einstein's equations 
to determine the  residual component $H_0(u, r)$. This versatile ansatz indeed works and finds 
\begin{align}
\label{}
H_0(u, r)=\frac{r U_1'(u)}{U_1(u)}+g^2 \left(r^2-\varepsilon \frac{U_0(u)}{U_1(u)}\right)\,.
\end{align}

Several comments are in order. A similar procedure to obtain $g\ne0$ solutions out of the $g=0$ solutions has been  
already worked out for the static case \cite{Faedo:2015jqa,Nozawa:2020gzz}.  
This strategy does not appear to work  for the class (I-b). 
Presumably, the prime rationale is that the class (I-b) solution does not allow a limit in which the 
scalar field vanishes, whilst 
the solution to the theory (\ref{pot:superpotential}) should admit the $\phi=0$ truncation.
Nevertheless, the class (I-b) solution is utilized to derive a new solution with a different scalar potential, 
as will be explained in appendix \ref{sec:classIb}.

To wrap up, a new Robinson-Trautman family of solutions in gauged supergravity 
reads 
\begin{align}
\label{RTmetric2}
\D s^2=-2 \D u \D r -H(u,r, z, \bar z) \D u^2+2 \frac{S(u,r)^2 }{P(u,z, \bar z)^2}\D z \D \bar z \,, 
\end{align}
and 
\begin{align}
\label{}
\phi(u, r)=& \, 
\left\{
\begin{array}{cc}
\dfrac 1{\sqrt 2} \ln \left(\dfrac{\sqrt{U_1(u)}r-\sqrt{U_0(u)}}{\sqrt{U_1(u)}r+\sqrt{U_0(u)}}\right)   &  (\varepsilon=+1)   \\
 \sqrt{2}\left[\dfrac \pi 2 -\arctan \left(\sqrt{\dfrac{U_1(u)}{U_0(u)}}r \right)\right]     &   (\varepsilon=-1) 
\end{array}
\right.
\,. \label{phi}
\end{align}
Here, the metric functions are given by 
\begin{subequations}
\label{RTsol}
\begin{align}
H(u,r,z,\bar z)=& \,\frac{K(u,z, \bar z)}{U_1(u)}+r \big[(\ln U_1(u))'-2\partial_u\ln P(u, z, \bar z) \big]+g^2 \left(r^2-\varepsilon \frac{U_0(u)}{U_1(u)}\right) \,,\label{f} \\
S(u,r)=&\, \sqrt{U_1(u) r^2-\varepsilon U_0 (u) }\,,\label{S} 
\end{align}
\end{subequations}
where $K$ is given by (\ref{K}) and obeys
\begin{align}
\label{Keq}
\Delta_2 K+2\varepsilon P^2 \sqrt{U_0U_1} \partial_u \left[\sqrt{\frac{U_1}{U_0}}\partial_u \left(\frac{U_0}{P^2}\right)\right]=0\,.
\end{align}
Note that this equation is the same as the $g=0$ case and one can always choose the gauge $U_1=1$ and $U_0=r_0^2$. 
Equation (\ref{Keq}) can be viewed as a fourth-order nonlinear differential equation for 
$P(u,z,\bar z)$, corresponding to the Robinson-Trautman equation (\ref{RTeqvacuum}) for the vacuum case. 
As illustrated in section \ref{sec:IaRF},  any solutions to the Ricci flow equation give rise to 
solutions to (\ref{Keq}) for $\varepsilon=1$. The only difference of the metric form from the massless scalar case is the last term in 
(\ref{f}). All physical conditions $\kappa=\sigma=\Psi_0=\Psi_1=\Phi_{01}=\Phi_{02}=\lambda=\pi=0$ together with gauge conditions 
$\epsilon=\tau=0$ imposed in the $g=0$ case are carried over to $g\ne 0$.

The null geodesics generated by $l=\partial/\partial r$ are  shear-free and twist-free, with $r$ being an affine parameter. 
Employing the null tetrads as
\begin{align}
\label{}
l =\frac{\partial}{\partial r}\,, \qquad 
n=\frac{\partial}{\partial u}-\frac 12 H\frac{\partial}{\partial r}\,, \qquad 
m=\frac{P}{S}\frac{\partial}{\partial z}\,, 
\end{align}
the convergence $\rho$ of $l^\mu$ reads
\begin{align}
\label{}
\rho=-\frac{\partial_r S}{S}=-\frac{U_1(u)r}{U_1(u)r^2+U_0(u)}  \,.
\end{align}

The surviving Weyl scalars are given by
\begin{subequations}
\label{Weylscalars}
\begin{align}
\Psi_2\equiv&\, C_{\mu\nu\rho\sigma}l^\mu m^\nu \bar m^\rho n^\sigma 
=\varepsilon\frac{U_0 K+r U_1 P^2 \partial_u (U_0 P^{-2})}{3S^4}\,, \\
\Psi_3\equiv&\, C_{\mu\nu\rho\sigma}n^\mu l^\nu n^\rho \bar m^\sigma=
\frac{P}{2S^3} 
\partial_{\bar z}\left(2{\varepsilon} U_0 \partial_u\ln P-rK\right)\,, \\
\Psi_4\equiv&\,C_{\mu\nu\rho\sigma}n^\mu \bar m^\nu n^\rho \bar m^\sigma=\frac{1}{2U_1S^2}\Big[\partial_{\bar z} \left(P^2 \partial_{\bar z}K\right)-2r P^2 U_1\partial_u 
\left(P^{-1}\partial_{\bar z}^2P\right)\Big]\,. 
\end{align}
\end{subequations} 
It turns out that the solution belongs to Petrov type II or more degenerate classes.

\subsection{Comparison with literature}

As commented previously, the special case $\partial_u P=0$ in the $g=0$ solution has been 
obtained in \cite{Tahamtan:2015sra,Tahamtan:2016fur}. 
Here, it is illustrative to make a comparison with other solutions in the literature.

In  \cite{Lu:2014ida,Lu:2014sza}, L\"u and V\'azquez-Poritz have 
constructed a charged and  dynamic C-metric in the present theory (\ref{Lag}), (\ref{pot:superpotential}) 
with $\varepsilon=+1$. In the neutral limit, their solution is 
\begin{align}
\label{}
 \D s^2=&\, \frac{1}{(1+\alpha(u) \hat r \hat x)^2 }\left[
 (1+q_0^2 \alpha (u)^2 \hat x)\left(-2 \D u \D \hat r-\hat H \D u^2\right)+\hat r^2 \left(1-\frac{q_0^2\alpha (u)}{\hat r}\right)
 \left(\frac{\D \hat x^2}{1-\hat x^2}+(1-\hat x^2)\D \varphi^2\right)
 \right] \,, \\
 \phi=&\frac 1{\sqrt 2}\ln \left(\frac{1+q_0^2\alpha(u)^2 \hat x}{1-q_0^2\alpha(u)/\hat r}\right)\,,
\end{align}
where 
\begin{align}
\label{}
\hat H(u, \hat r, \hat x)=&\, 1-\alpha(u)^2 \hat r^2+g^2 \hat r^2 \left(1-\frac{q_0^2\alpha(u)}{\hat r}\right)\notag \\
&
-\frac{\alpha'(u)}{(1+q_0^2\alpha(u)^2 \hat x)^2}
\left[q_0^2+2\hat r^2 \hat x+q_0^2 \alpha(u)^2 \hat r \left(3\hat r-2q_0^2 \alpha (u)\right)\hat x^2\right]\,.
\end{align}
The function $\alpha =\alpha (u)$ is subjected to $\alpha'(u)=(-1+q_0^4\alpha(u)^4)/q_0^2$. 
It can be verified that the Petrov type of this solution is II. 
This family of solutions is indeed incorporated in our solution. 
The explicit transformations are given by 
\begin{align}
\label{}
r=\frac{\hat r-(q_0^2/2)\alpha(u)(1-\alpha(u)\hat r\hat x)}{1+\alpha (u)\hat r\hat x } \,, \qquad 
 x=\frac 12 \ln \left(\frac{1+\hat x}{1-\hat x}\right)-\int \alpha(u) \D u \,,
 \qquad 
 z=\frac 1{\sqrt 2}(x+i\varphi) \,, 
\end{align}
with 
\begin{align}
\label{}
P=\sqrt{\frac{1+q_0^2 \alpha(u)^2 \hat x}{1-\hat x^2}} \,, \qquad 
U_1(u)=1 \,, \qquad U_0(u) =\frac 14 q_0^4 \alpha(u)^2 \,.
\end{align}

\subsection{Petrov III and N solutions}

Let us next investigate the case where the Petrov type for the solution 
(\ref{RTsol}) is degenerate into type III. 
The type III of Petrov classification is characterized by 
the vanishing of $I \equiv \Psi_{ABCD}\Psi^{ABCD}$
and $J\equiv -\Psi_{AB}{}^{CD}\Psi_{CD}{}^{EF}\Psi_{EF}{}^{AB}$. 
For the present Robinson-Trautman case, this amounts to $\Psi_2=0$. 
We suppose $U_0\ne 0$: otherwise the scalar field becomes trivial. 
It follows from (\ref{Weylscalars}) that $K=0$ and $\partial_u (U_0P^{-2})=0$.
The freedom of two-dimensional conformal transformation $z\to \ti z(z)$ then enables us to 
set $P=U_0(u)^{1/2}$. 
In this case, all the Weyl scalars become zero, i.e., the spacetime is conformally flat.
Specifically, the present family of solutions does not incorporate the pure Petrov III class. 

Relabeling  $r=g\sqrt{U_0/U_1}\ti r$ and 
$\D  u=(g\sqrt{U_0/U_1})^{-1}\D \ti u$, the conformally flat solution is reduced to the static form
\begin{align}
\D s^2=&-2 \D \ti u\D \ti r-\left(-\varepsilon+g^2\ti r^2\right)\D \ti u^2 +2\left(-\varepsilon+g^2\ti r^2\right)\D z \D \bar z \,, 
\notag \\
\phi=&\, 
\left\{
\begin{array}{lc}
\dfrac 1{\sqrt 2}\ln \left(\dfrac{g\ti r-1}{g\ti r+1}\right)\,, & (\varepsilon=+1) \\ 
{\sqrt 2}\left(\dfrac \pi 2-\arctan (g\ti r)\right)\,, & (\varepsilon=-1)
\end{array}
\right. \,.
\label{CFsol}
\end{align}
Further transformations $\D \ti u=\D\ti t-\D \ti r/(-\varepsilon+g^2 \ti r^2)$, 
$z=(x+i y)/\sqrt 2$ give
\begin{align}
\label{}
\D s^2=&\frac{\D \ti r^2}{-\varepsilon+g^2\ti r^2} +\left(-\varepsilon+g^2\ti r^2\right)\left(-\D \ti t\,^2+\D x^2+\D y^2\right)\,.
\end{align}
For $\varepsilon=-1$, 
this solution describes a plane-symmetric wormhole in AdS, connecting the vacuum $\phi=0$ at $\ti r\to \infty$
and $\phi=\sqrt 2\pi =\phi_1$ at $\ti r\to -\infty$.

In a parallel fashion, we can show that there exist no pure type N solutions with $\Psi_{(AB}{}^{EF}\Psi_{CD)EF}=0$.

\subsection{Petrov D solutions}
\label{sec:PetrovD}

We next discuss the case where the spacetime falls into Petrov type D.  In terms of Weyl spinors, this requires 
$\Psi_{PQR(A}\Psi_{BC}{}^{PQ}\Psi^R{}_{DEF)}=0$ \cite{Stewart:1990uf}. In the present 
Robinson-Trautman case with $\Psi_0=\Psi_1=0$, this is only possible for 
\begin{align}
\label{PetrovD}
2 \Psi_3^2-3\Psi_2 \Psi_4=0 \,, \qquad \Psi_2 \ne 0 \,.  
\end{align}
Dividing this condition into different powers of $r$, 
we obtain three independent conditions $ D_{(i)}=0$ ($i=0,1,2$), where 
\begin{subequations}
\label{D012}
\begin{align}
D_{(0)}=&\,K\partial_{\bar z}(P^2 \partial_{\bar z}K)-{\varepsilon}\frac{U_1U_0}{P^{2}}L^2 \,, \\
D_{(1)}=&\,\partial_u  \left(\frac{U_0}{P^{2}}\right)\partial_{\bar z}(P^2 \partial_{\bar z}K)-\frac{U_0}{P^{2}} (K\partial_{\bar z}L-2 L \partial_{\bar z}K)\,, \\
D_{(2)}=&\, (\partial_{\bar z}K)^2+{\varepsilon}U_1 \partial_u \left(\frac{U_0}{P^{2}}\right)\partial_{\bar z}L \,.
\end{align}
\end{subequations}
Here, we have defined
\begin{align}
\label{}
 L (u, z, \bar z)\equiv 2 P^2 \partial_u \partial_{\bar z} \ln P \,. 
\end{align}

In the upcoming subsections, we shall explore the two cases separately, 
according to $\partial_{\bar z}K=0$ or $\partial_{\bar z}K\ne 0$.

\subsubsection{$\partial_{\bar z}K\ne 0$ case}

Defining 
$\ma D \equiv P^2 \partial_{\bar z}L [\partial_u (U_0P^{-2})D_{(0)}-KD_{(1)}]+U_0 L^2 D_{(2)}$, we have 
\begin{align}
\label{}
\ma D
=U_0 K^4 \left[\partial_{\bar z}\left(\frac{L}{K}\right)\right]^2\,.
\end{align}
Upon integration of $\ma D=0$, we get 
$K= d_0(u, z)L$, where $d_0(\ne 0)$ is a complex-valued function. 
Now, $D_{(0)}=D_{(1)}=0$ reduce respectively to
\begin{align}
\label{Lzbeq1}
\partial_{\bar z}\left(d_0 P^2 \partial_{\bar z}K-2{\varepsilon}U_1U_0\partial_u \ln P\right)=&\, 0\,, 
\\
\label{Lzbeq2}
U_0 P^{-2} K \partial_{\bar z}K+\partial_u \left(\frac{U_0}{P^{2}}\right)\partial_{\bar z}\left(d_0P^2 \partial_{\bar z}K\right)=&\, 0\,.
\end{align}
Equation (\ref{Lzbeq1}) implies the existence of a complex-valued function $d_1(u,z)$ such that 
\begin{align}
\label{d1}
d_0 P^2 \partial_{\bar z}K-2{\varepsilon}U_1U_0\partial_u \ln P=d_1\,.
\end{align}
Eliminating $\partial_{\bar z}K$ from (\ref{Lzbeq2}) using (\ref{d1}), 
we end up with 
\begin{align}
\label{d1s}
d_1 =d_1(u)=-{\varepsilon} U_1(u)U_0'(u) \,. 
\end{align}
Combined with (\ref{d1s}), 
 equation (\ref{d1}) is cast into
 \begin{align}
\label{Kzb}
\partial_{\bar z}K= -{\varepsilon} \frac{U_1}{d_0}
 \partial_u \left(\frac{U_0}{P^{2}}\right)\,.
\end{align}
The $u$-derivative of $K=2 P^2 \partial_z\partial_{\bar z}\ln P$  yields
\begin{align}
\label{Ku}
\partial_u K =\frac{K}{d_0}\partial_z \left[\ln \left(\frac{K}{d_0P^2}\right)\right]+2K \partial_u \ln P \,. 
\end{align}
In conjunction with the scalar Robinson-Trautman equation (\ref{Keq}), 
the integrability condition $(\partial_{\bar z}\partial_u -\partial_u\partial_{\bar z})K=0$ of these equations 
calls for 
\begin{align}
\label{Kint}
d_0 \partial_u \left[\ln \left(\frac{U_0U_1}{P^4 d_0^2 }\right)\right]-\partial_z \left[\ln \left(\frac{U_0U_1}{P^4 d_0^2 }\right)\right]=0 \,. 
\end{align}
From $K-d_0 L=2P^2 \partial_{\bar z}(\partial_z\ln P-d_0 \partial_u\ln P)=0$, 
we have 
\begin{align}
\label{d2}
d_0(u,z)\partial _u \ln P-\partial_z \ln P=d_2(u,z)\,,
\end{align}
thereby equation (\ref{Kint}) requires 
\begin{align}
\label{d2s}
d_2 (u,z)=\frac{d_0}{4} \left[\partial_u \ln \left(\frac{U_1U_0}{d_0^2}\right)-\partial_z \left(\frac{2}{d_0}\right)\right]\,.
\end{align}
Substituting the $z$ derivative of (\ref{Kzb}) into (\ref{Keq}), and eliminating $\partial_z P$ 
by (\ref{d2}) and (\ref{d2s}), $d_0(u,z)$ should satisfy
\begin{align}
\label{d0eq}
\sqrt{\frac{U_1}{U_0}}\partial_u\left(\sqrt{\frac{U_0}{U_1}}d_0\right)-\partial_z \ln d_0=d_3'(z) \,,
\end{align}
where $d_3=d_3(z)$ is a function of $z$. We have defined the right-hand side of the above equation 
by the $z$-derivative of $d_3(z)$ for later convenience. 
Equation (\ref{d2}) is now reduced to 
\begin{align}
\label{}
\label{Pzg}
\partial_z\left( \ln P\right)-\frac {1}{2} d_3'(z)=& -\frac 12 d_0\partial_u \ln \left(\frac{U_0}{P^{2}}\right)\,. 
\end{align}
Combined with the complex conjugation of (\ref{Pzg}), 
the compatibility condition $(\partial_z\partial_{\bar z}-\partial_{\bar z}\partial_{z}) \ln P=0$ 
is cast into 
\begin{align}
\label{}
\partial_{z}\left(\frac 1{d_0(u,z)}\right)-\frac{d_3'(z)}{d_0(u, z)}=\partial_{\bar z}\left(\frac 1{\bar d_0(u, \bar z)}\right)-\frac{\bar d_3'(\bar z)}{\bar d_0(u, \bar z)}\,, 
\end{align}
where we have used $\partial_u (U_0/P^2)\ne 0$ by virtue of (\ref{Kzb}). 
Inspecting $\partial_z[\partial_z (d_0^{-1})-d_3'(z) d_0^{-1}]=0$ and (\ref{d0eq}), it turns out that $d_0$ must be of the factorized form
\begin{align}
\label{}
d_0(u,z)=  \sqrt{\frac{U_1(u)}{U_0(u)}}e^{-d_3(z)}\,.
\end{align}
Then, equation (\ref{Pzg}) is simplified to
\begin{align}
\label{eqP}
e^{d_3(z)} \partial_z \left[\ln \left(\frac{U_0(u)}{P^2}e^{d_3(z)}\right)\right]
=\sqrt{\frac{U_1(u)}{U_0(u)}}\partial_u\left[\ln \left(\frac{U_0(u)}{P^2}e^{d_3(z)}\right)\right]\,.
\end{align}
Using the reparametrization freedom (\ref{w1w2}), one can employ the gauge
$U_0(u)=m^2 U_1(u)$ with $m$ being a constant and $\D \ti z/\D z \propto e^{d_3(z)}$. 
This amounts to setting $d_3(z)$ to be a constant.

Writing 
$d_0=1/(\sqrt 2A)$ (i.e., $e^{d_3}=\sqrt 2 A/m$) with a real constant $A$, it follows from (\ref{eqP}) that 
$\ma P\equiv U_1(u)/P(u, z,\bar z)^2$ turns out to satisfy 
\begin{align}
\label{}
\partial_z \ma P=\frac 1{\sqrt 2 A }\partial_u \ma P \,. 
\end{align}
Considering the complex conjugation of above equation, we conclude that $\ma P$ depends only a single variable 
as $\ma P=\ma P(s)$, where $s\equiv u+(z+\bar z)/(\sqrt 2 A)$. 
Equation (\ref{Kzb}) now reads
\begin{align}
\label{}
\frac{\D }{\D s}\left(\ma P^{-1} \frac{\D^2 }{\D s^2} \ln \ma P-4{\varepsilon }A^4m^2 \ma P\right)=0 \,. 
\end{align}
Defining a new variable $\eta$ by $\D \eta/\D s=\ma P(s)$, 
the above equation is simplified to 
\begin{align}
\label{}
\frac{\D ^2}{\D \eta^2}\ma P=4{\varepsilon }A^4m^2(\ma P-c_0) \,, 
\end{align}
thus 
\begin{align}
\label{}
\ma P(\eta)=c_0 +c_1\cosh \left(2{\sqrt{\varepsilon }}A^2m \eta\right)+\frac{c_2}{{\sqrt{\varepsilon }}} 
\sinh \left(2{\sqrt{\varepsilon }}A^2m\eta\right) \,,
\end{align}
and $K=-U_1(u)\ma P''(\eta)/(2A^2)$. 
Here, $c_0$, $c_1$ and $c_2$ are real constants. 
A change of variables $z=(w+i\varphi)/\sqrt 2$, $\D w=A(\D s-\D u)=A(\D \eta/\ma P(\eta)-\D u)$ 
yields 
\begin{align}
\label{}
\D s^2=&\,-2 \D u \D r+ (r^2{-\varepsilon}m^2)\left(-2 A^2 \D u \D \eta +\frac{A^2 \D \eta^2}{\ma P(\eta)}+\ma P(\eta)\D \varphi^2 
\right)\notag \\
&+\left[-2{\varepsilon}A^2m^2 c_0-g^2 (r^2{-\varepsilon}m^2)+A^2(r^2+{\varepsilon}m^2)\ma P(\eta)-r \ma P'(\eta)\right]\D u^2
\,. 
\end{align}
In either sign of $\varepsilon$, this solution agrees with the C-metric.

Indeed, for $\varepsilon=+1$,  by defining 
\begin{align}
\label{}
r=\frac{2+Am(x+y)}{A(x-y)}\,, \qquad 
\cosh (2A^2 m\eta)=\frac{5+8Am x+4A^2m^2x^2}{4(1+Am x)}\,, \qquad 
A\D u =\D t-\frac 1{2\Delta_y(y)}\D y \,, 
\end{align}
and  
\begin{align}
\label{}
\Delta_x(x)=&\,c_0+\frac 54c_1+\frac34 c_2 +\left(c_0+2c_1+2c_2\right)A m x +(c_1+c_2)A^2m^2 x^2 \,,\notag \\
\Delta_y(y)=&\,-c_0-\frac 54c_1-\frac34 c_2 -\left(c_0+2c_1+2c_2\right)A m y-(c_1+c_2)A^2m^2 y^2 +\frac{g^2}{A^2}(1+Am y) \,,
\end{align}
the solution  reduces to the one  \cite{Lu:2014ida,Lu:2014sza,NT1}
\begin{align}
\label{}
\D s^2 =&\, \frac{1}{A^2 (x-y)^2} \left[(1+Amx)\left(- \Delta_y(y)\D \hat t^2+\frac{\D y^2}{\Delta_y(y)}\right)
+(1+Am y)\left(\frac{\D x^2}{\Delta_x(x)}+ \Delta_x(x) \D \hat \varphi^2 \right) 
\right]\,,  \notag \\
\phi=&\frac 1{\sqrt 2} \ln \left(\frac{1+Am y}{1+Am x}\right)\,,
\label{hairedCmetric}
\end{align}
where $\hat t=2t$ and $\hat \varphi=2\varphi$. 
In appearance, this solution might represent a scalar haired C-metric. However, a careful analysis 
demonstrates that the solution (\ref{hairedCmetric}) does not represent a pair of accelerated black holes
with a scalar hair, since the spacetime inevitably allows a naked singularity \cite{NT1}. 
The naked singularity can be avoided if the solution is charged.

For $\varepsilon=-1$, 
the following redefinitions
\begin{align}
\label{}
r=&\, \frac{1+A^2m^2 xy}{A(x-y)}\,, \qquad \cos (2A^2m\eta) =\frac{1-A^2m^2x^2}{1+A^2m^2 x^2} \,, \qquad A \D u=\D t-\frac{1}{\Delta_y(y)}\D y \,, 
\end{align}
with 
\begin{align}
\label{}
\Delta_x (x)=&\, (c_0+c_1)+2c_2 A m x+A^2m^2 (c_0-c_1)x^2 \,, \notag \\
\Delta_y(y)=&\, -(c_0+c_1)-2c_2 A m y-A^2 m^2 (c_0-c_1)y^2+ g^2 (A^{-2}+m^2 y^2) \,, 
\end{align}
give the solution in  \cite{NT2} as 
\begin{align}
\label{WHCmetric}
\D s^2 =&\, \frac{1}{A^2 (x-y)^2} \left[(1+A^2m^2 x^2)\left(- \Delta_y(y)\D t^2+\frac{\D y^2}{\Delta_y(y)}\right)
+(1+A^2m^2 y^2)\left(\frac{\D x^2}{\Delta_x(x)}+ \Delta_x(x) \D \varphi^2 \right) 
\right]\,,  \notag \\
\phi=&\sqrt 2 \left[\frac \pi 2 -\arctan \left(\frac{1+A^2m^2xy}{Am (x-y)}\right)\right]\, 
\end{align}
In the limit where the acceleration parameter $A$ vanishes, the solution (\ref{WHCmetric}) reduces to the 
wormhole in AdS \cite{Nozawa:2020gzz}, which is the generalization of 
the asymptotically flat Ellis-Bronnikov wormhole \cite{Ellis:1973yv,Bronnikov:1973fh} (see also
\cite{Martinez:2020hjm,Nozawa:2020wet}).  
This solution (\ref{WHCmetric}) therefore describes accelerated wormholes in AdS \cite{NT2}. 
A distinguished property of wormhole C-metric in this theory is that 
the conical singularity along the symmetry axis can be completely resolved, for which 
the acceleration of wormholes is provided solely by a phantom field.

\subsubsection{$\partial_{\bar z}K=0$ case}

The classification in this class has been done in \cite{Podolsky:2016sff}. 
Assuming $\partial_{\bar z}K=0$, we see that 
$K$ is a function of $u$ only and $D_{(0)}=0$ gives 
\begin{align}
\partial_u \partial_{\bar z} \ln P=0\,.
\end{align}
Combined with the complex conjugation of the above equation, 
 we obtain $P=P_0(u)P_1(z, \bar z)$, where $P_0$ and $P_1$ are real functions. 
In this case, $D_{(1)}=D_{(2)}=0$ follows automatically, implying $\Psi_3=\Psi_4=0$. 
From the definition (\ref{K}) of $K$, we have 
$K=2P_0^2 P_1^2 \partial_z\partial_{\bar z} \ln P_1$. Since $\partial_{\bar z} K=0$, this implies 
$K=P_0^2(u)k$ and $2\partial_z\partial_{\bar z}\ln P_1=k$,  where $k$ is a constant which can be  normalized as 
$k=0, \pm 1$.  It follows that 
$2 P_1^{-2}\D z\D \bar z$ is a $u$-independent maximally symmetric space with a constant Gauss curvature 
$k$. 
Namely, $P_1$ can be chosen as $P_1=1+(k/2) z \bar z$. 
Exploiting the remaining gauge freedom (\ref{w1w2}), one can choose $U_1(u)=P_0(u)^2$, for which 
equation (\ref{Keq}) implies $U_0=(\omega u+r_0)^2P_0(u)^2$,  where $\omega$ and $r_0$ are constants.

To recap, the Petrov D solution with $\partial_{\bar z}K=0$ is free from $P_0(u)$ and is given by 
\begin{align}
\label{petrovDsym}
\D s^2=&\, -2 \D u\D r -\Big\{k+g^2\left[r^2-\varepsilon (\omega u+r_0)^2\right] \Big\}\D u^2
+\left[r^2 -\varepsilon (\omega u+r_0)^2\right]\D \Sigma_k^2 \,, \\
\phi=&\, \left\{
\begin{array}{ll}
 \dfrac 1{\sqrt 2}\ln \left(\dfrac{r-r_0-\omega u}{r+r_0+\omega u}\right)     &     (\varepsilon =+1 ) \\
\sqrt 2 \left[\dfrac{\pi}{2}-\arctan \left(\dfrac{r}{\omega u+r_0}\right)\right]      &   (\varepsilon =-1 )
\end{array}
\right.  \,,
\end{align}
where $\D \Sigma_k^2$ is a space of constant Gauss curvature given in (\ref{dSigma}). 
This solution  admits Killing vectors generating ${\rm SO}(3)$, 
${\rm ISO}(2)$, ${\rm SO}(1,2)$ symmetry
for  $k=1, 0, -1$, respectively. 
The nonvanishing Weyl scalar is 
 \begin{align}
\label{}
\Psi_2=\varepsilon \frac{(\omega u+r_0)[2\omega r+k(\omega u+r_0)]}{3[r^2-\varepsilon (\omega u+r_0)^2]^2}\,.
\end{align}

For the $\varepsilon=+1$ case, the present solution (\ref{petrovDsym}) agrees with 
the solution (29) with $\alpha_{\rm there}=0$ in \cite{Fan:2016yqv}.

\section{Physical properties of Petrov D solutions with constant curvature base space}
 \label{sec:D}

The Petrov D solution (\ref{petrovDsym}) derived in the previous section 
is simple enough, but gives us a number of insightful implications. 
This section is therefore intended to explore the physical properties of the solution (\ref{petrovDsym}). 
For definiteness of the argument, we confine ourselves to the case where $\omega \ge 0$, $r_0\ge 0$ and $g>0$.

 \subsection{$\omega =0$ case}
 \label{Dstatic}
 
 Setting $\omega=0$, we obtain the static solution. 
 Writing $\Delta (r)=k+ g^2(r^2-\varepsilon r_0^2)$, 
the solution (\ref{petrovDsym}) is simplified to 
 \begin{align}
\label{}
\D s^2=&\, -\Delta (r) \D t^2+\frac{\D r^2}{\Delta(r)}+\left(r^2 -\varepsilon r_0^2\right)\D \Sigma_k^2 \,, \\
\phi=&\, \left\{
\begin{array}{lc}
  \dfrac 1{\sqrt 2}\ln \left(\dfrac{r-r_0}{r+r_0}\right)    &    (\varepsilon=+1)\\
 \sqrt 2 \left[\dfrac{\pi}{2}-\arctan \left(\dfrac{r}{r_0}\right)\right]     &   (\varepsilon=-1)
\end{array}
\right.\,,
\label{PetrovDstatic}
\end{align}
 where we have set  $u=t-\int \D r /\Delta(r)$. 
 
 For $\varepsilon=+1$ with $k=1, 0$, the solution represents an asymptotically AdS spacetime with a 
 timelike naked singularity at $r=r_0$ \cite{Faedo:2015jqa}. 
  For $\varepsilon=+1$ with $k=-1$, the spacetime describes a topological black hole 
  in AdS with a regular event horizon.

For $\varepsilon=-1$ and  $k=+1, 0$,  the solution represents a static wormhole in AdS \cite{Nozawa:2020gzz}.  
The $k=0$ solution is conformally flat and recovers (\ref{CFsol}). 
The solution with $\varepsilon=k=-1$ describes a wormhole for $gr_0>1$ and a black hole for $gr_0\le 1$. 
The wormhole throat exists at $r=0$, where the areal radius $\ma R=\sqrt{r^2+r_0^2}$
takes the minimal value $\ma R_{\rm min}=r_0$.

The spacetime is asymptotically AdS for $r\to  \infty$ 
(and for $r\to -\infty$ as well in the $\varepsilon=-1$ case), around which 
the metric is approximated as 
\begin{subequations}
\label{asy}
\begin{align}
\D s^2\simeq &\,-\left(k+g^2 \ma R^2+O(1/\ma R^2)\right)\D t^2 +\frac{\D\ma  R^2}{k+\varepsilon g^2r_0^2+g^2 \ma R^2+O(1/\ma R^2)}+\ma R^2 \D\Sigma_k^2 \,, 
\\
\phi\simeq &\, -\frac{\sqrt 2 \varepsilon r_0}{\ma R}+O(1/\ma R^3)\,,
\end{align}
\end{subequations}
The asymptotic behavior of the metric (\ref{asy}) reveals that the global mass vanishes for $\omega =0$, and the 
scalar field obeys the Neumann boundary condition.

It is worthy of remark that the spacetime approaches to different vacua as $r\to \pm \infty$ for $\varepsilon=-1$.
For $r\to \infty$, we have $\phi\to 0$, whereas for $r\to -\infty$, $\phi\to \sqrt 2\pi =\phi_{+1}$ is attained. 
This is a salient feature reminiscent of solitons.

 \subsection{Mass function}
 
 We define the areal radius by
\begin{align}
\label{radius}
\ma R=\sqrt{r^2-\varepsilon (\omega u+r_0)^2} \,.
\end{align}
In terms of $\ma R$, the 
Misner-Sharp quasi-local mass is defined by \cite{Misner:1964je,Hayward:1994bu,Maeda:2007uu,Ohashi:2015xaa}
\begin{align}
\label{}
M=\frac{\ma R}{2} \left(k-(\nabla \ma R)^2+g^2 \ma R^2\right) \,.
\end{align}
It has been known that the quasi-local mass introduced by Hawking \cite{Hawking:1968qt} reduces to the Misner-Sharp quasi-local mass, 
provided the spacetime admits a base space with ${\rm SO}(3)$, ${\rm ISO}(2)$ or ${\rm SO}(1,2)$ symmetry \cite{Hayward:1994bu}. 
Explicitly, 
\begin{align}
\label{MSD}
M=-\frac{\varepsilon(\omega u+r_0)}{2\sqrt{r^2-\varepsilon (\omega u+r_0)^2}}\Big(
2\omega r+k(\omega u+r_0)+g^2(\omega u+r_0)\left[r^2-\varepsilon (\omega u+r_0)^2\right]
\Big)\,.
\end{align}
It can be confirmed that the sign of $M$ is related to the causal nature of the vector 
$\nabla_\mu \phi$ as
\begin{align}
\label{}
M=-\frac{\varepsilon}4 \ma R^3(\nabla \phi)^2 \,. 
\end{align}
In the $r\to \infty$ limit, we can verify that the Misner-Sharp quasi-local mass (\ref{MSD}) diverges. 
This divergence can be attributed to the fact that the decay rate of spacetime at AdS infinity 
is sufficiently slow as indicated by (\ref{asy}), which 
traces back to the mass spectrum at the AdS vacuum (\ref{AdSradii}). 
Consequently, the Misner-Sharp quasi-local mass is not an appropriate indicator to extract the physical mass
in the present setting. 

A simple and honest way to find the global mass function is to take 
the asymptotic limit of the metric (\ref{petrovDsym}) as 
\begin{align}
\label{}
\D s^2\simeq -2  \D u\D \ma R-\left(k+g^2\ma  R^2-\frac{2M_{\rm B}}{\ma R}\right) \D u^2 +\ma R^2 \D \Sigma_{k}^2 \,,
\end{align}
where
\begin{align}
\label{}
M_{\rm B}(u)=-\varepsilon \omega (\omega u+r_0) \,. 
\end{align}
 It seems reasonable to regard $M_{\rm B}(u)$ as the analogue of the Bondi mass, since 
 the metric takes the Vaidya form.\footnote{
In the case of the positive cosmological constant, the Bondi mass has been discussed by various authors. 
 See e.g., \cite{Fernandez-Alvarez:2021yog,Bonga:2023eml} and references therein. }
Furthermore, it is compelling  that $M_{\rm B}$ vanishes for the static case ($\omega=0$), 
 reproducing the previous result (\ref{asy}).
The mass loss rate yields
\begin{align}
\label{}
\frac{\D M_{\rm B}(u)}{\D u}=-\varepsilon \omega^2 \,, 
\end{align}
which is negative as it should be for the $\varepsilon=+1$ case. 
In the phantom case $\varepsilon=-1$,  the mass is growing, which is 
attributed to the  violation of the energy conditions.

\subsection{$u\to \pm \infty$ limit}
 
For $\varepsilon=+1$, the allowed region of $u$ is bounded for finite $r$, 
on account of the restriction $S(u,r)\ge 0$. For $\varepsilon=-1$, 
the $u\to \pm \infty$ limit with finite $r$ leads to 
\begin{align}
\label{asyuinf}
\D s^2\simeq & \, -2 \D u \D r -(g\omega u)^2 \D u^2+(\omega u)^2  \D \Sigma_k^2 
 = -2 \D U \D V+U^2 \D \Sigma_k^2 \,, 
\end{align}
where $U=\omega u$ and 
$\D r=\omega \D V-(g^2U^2 \D U)/(2\omega)$.
For $k=0$, the metric (\ref{asyuinf}) represents the Minkowski spacetime
written in the Rosen coordinates. 
In this limit, the scalar field in the $r>0$ region tends to 
$\phi \to 0$ for $u\to \infty$
and  
$\phi \to \pi/\sqrt 2=\ti \phi_{+1}$  for $u\to -\infty$.

 \subsection{Singularities}

 Let us next explore the spacetime singularities. 
 For the definiteness of the argument, we confine to the 
 spherically symmetric case $k=1$. Furthermore, 
$r_0$ can be eliminated by $u\to u-\omega/r_0$ in the dynamical case $\omega>0$. 
For the static case, see section \ref{Dstatic}. 
Here and in what follows,  we therefore set $k=1$ and $r_0=0$.

 The spacetime curvature singularities are characterized by the divergence of curvature invariants. For instance, 
the square of the Weyl tensor and the scalar curvature are given by 
\begin{align}
\label{}
C_{\mu\nu\rho\sigma}C^{\mu\nu\rho\sigma}=&\, 48\Psi_2^2=
\frac{16\omega^4 (2r+u)^2u^2}{3(r^2-\varepsilon \omega^2 u^2)^4}\,, 
\\
R=&\, \frac{2\varepsilon\omega^2 u(2r+u) -6 g^2 (2r^2-\varepsilon \omega^2 u^2)
(r^2-\varepsilon \omega^2 u^2 )}{(r^2-\varepsilon \omega^2 u^2)^2}\,.
\end{align}
Other curvature invariants behave similarly. 
For $\varepsilon=+1$, there appear curvature singularities at $u=u_{\rm s}^{(\pm)}$ where
\begin{align}
\label{}
u_{\rm s}^{(\pm)}=\pm \frac{r}{\omega }\,.
\end{align}
These are central singularities, in the sense that the areal radius (\ref{radius}) vanishes there. 
Along these singularities, 
\begin{align}
\label{}
\D s^2|_{u=u_{\rm s}^{(\pm)}}=-\left(1\pm 2\omega \right)\D u^2 |_{u= u_{\rm s}^{(\pm)}} \,.
\end{align}
Thus, the singularity $u=u_{\rm s}^{(+)}$ is always timelike, while the singularity $u=u_{\rm s}^{(-)}$
is timelike (spacelike) for $0<\omega <1/2$ ($\omega>1/2$), and null for $\omega=1/2$. 
The singularity $u=u_{\rm s}^{(-)}$ with finite $r$ can be accessed within a finite affine time for the outgoing null geodesics
along $l=\partial/\partial r$, since $r$ is an affine parameter. We have checked numerically that $u=u_{\rm s}^{(+)}$ is 
reached in a finite affine time for ingoing null geodesics.

Let us next investigate the $\varepsilon=-1$ case. 
The possible blowup of curvature invariants occurs for $k\ne 0$ and 
comes from the single point
\begin{align}
\label{sing}
r=0 \quad \textrm{with $u\to 0$} \,. 
\end{align}
Note that the divergence does not occur 
 along the null curve $u= 0$ with $r\to 0$ limit. 
 We remark that the coordinate system in \cite{Tahamtan:2016fur} is not maximally extended so that 
 the singularity (\ref{sing}) has been missing. 

Lastly, we make a succinct comment on the $k=0$ case. All the curvature invariants remain bounded at (\ref{sing}). However, this point is not completely 
regular. To demonstrate this, let us remind that the present null tetrads $\{l^\mu, n^\mu, m^\mu, \bar m^\mu\}$ define
the parallelly propagated  frame along $l^\mu$, see (\ref{ppframe}). A simple calculation finds 
\begin{align}
\label{}
R_{\mu\nu}l^\mu l^\nu =\frac{2\varepsilon \omega^2 u^2}{(r^2+\omega^2 u^2)^2} \,, 
\end{align}
which is indeed divergent at (\ref{sing}), irrespective of $k$. 
Since the left side of the above equation is the curvature components of the parallelly propagated frame (\ref{ppframe}), 
the point (\ref{sing}) is a parallelly propagated (p.p.) curvature singularity \cite{Hawking:1973uf} in any case.

\subsection{Trapping horizons}

The notion of trapping horizons was proposed by Hayward~\cite{Hayward:1993wb}, as a local characterization of 
dynamical black holes.  A trapping horizon is the closure of a hypersurface foliated by 
marginal surfaces. Specifically, the trapping horizons in spherical symmetry display a variety of physically 
desirable properties related to Misner-Sharp quasilocal mass and singularities~\cite{Hayward:1994bu,Nozawa:2007vq}. 

In terms of the areal radius, 
the trapping horizon \cite{Hayward:1994bu} is given by $(\nabla \ma R)^2 =0$, 
which boils down to 
\begin{align}
\label{}
(\nabla \ma R)^2=\frac{r\left[r+2\varepsilon \omega^2 u+g^2 r(r^2-\varepsilon \omega^2 u^2)\right]}{r^2-\varepsilon \omega^2 u^2}=0\,,
\end{align}
where we have set $k=1$ and $r_0=0$. 
The first kind trapping horizon at $r=0$, which survives in the static case, is identified as 
a throat of the wormhole for $\varepsilon=-1$. 
The second kind of trapping horizons is time-dependent 
\begin{align}
\label{}
u=u^{(\pm )}_{\rm TH} (r) := \frac{\omega \pm \sqrt{\omega^2+\varepsilon g^2r^2(1+g^2r^2)}}{g^2 \omega r}\,.
\end{align}

Let us first consider the case $\varepsilon=+1$, for which the allowed region $\ma R=\sqrt{r^2-\omega^2 u^2}>0 $ is 
$u_{\rm s}^{(-)}<u<u_{\rm s}^{(+)}$. 
It can be easily verified that $u^{(+ )}_{\rm TH}-u_{\rm s}^{(+)}>0$ is always fulfilled. 
This demonstrates that the  trajectory $u^{(+)}_{\rm TH}$ is of little significance to the present discussion, since the region $u>u_{\rm s}^{(+)}$ if not Lorentzian but has the signature $(+,-,-,-)$. 

For the trapping horizon $u^{(- )}_{\rm TH}$, 
we have
\begin{align}
\label{}
u^{(- )}_{\rm TH}-u_{\rm s}^{(-)}=\frac{(2\omega-1)r}{\omega (\sqrt{\omega^2+g^2r^2(1+g^2r^2)}+\omega+g^2r^2)}\,.
\end{align}
It is shown that the sign of $u^{(- )}_{\rm TH}-u_{\rm s}^{(-)}$ is controlled by $2\omega-1$. 
We see that the 
singularity at $u=u_{\rm s}^{(-)}$ is in the untrapped region $(\nabla \ma R)^2>0$ for $\omega<1/2$  and 
in the trapped region $(\nabla \ma R)^2<0$ for $\omega>1/2$.  
Along the trapping horizon $u=u^{(- )}_{\rm TH} (r)$,  we have
\begin{align}
\label{}
\D s^2|_{u=u^{(- )}_{\rm TH}}=& \frac{2(\Xi_1-\Xi_2)}{g^4r^4 (1+g^2r^2)[g^2r^2(1+g^2r^2)+\omega^2]}\left[(2+g^{-2}r^{-2})\omega+\sqrt{g^2r^2(1+g^2r^2)+\omega^2}\right]
\D r^2\,,
\end{align}
where 
$\Xi_1=\omega (3g^2r^2+2g^4r^4+4\omega^2)$ and 
$\Xi_2=(g^2r^2+4\omega^2)\sqrt{g^2r^2(1+g^2r^2)+\omega^2}$. 
One sees that the sign of $\D s^2|_{u=u^{(- )}_{\rm TH}}$ is controlled by $\Xi_1-\Xi_2$. 
In view of
\begin{align}
\label{}
\Xi_1^2-\Xi_2^2= g^{6}r^{6}(1+g^2r^2)(4\omega^2-1) \,, 
\end{align}
trapping horizon $u=u^{(- )}_{\rm TH}$ is spacelike for $\omega>1/2$, timelike for $0<\omega<1/2$ and null for $\omega=1/2$. 
This is consistent with the fact that the singularity with the positive (negative) Misner-Sharp quasi-local mass is spacelike (timelike) and lies 
in the trapped (untrapped) region \cite{Hayward:1994bu}.

\begin{figure}[t]
\begin{center}
\includegraphics[width=14cm]{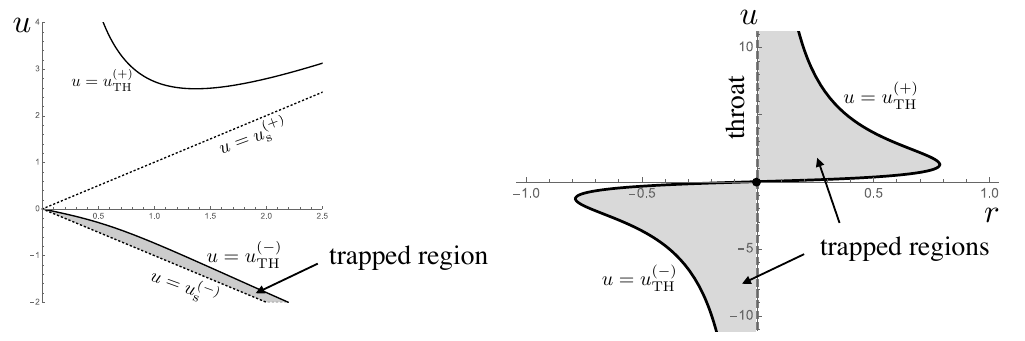}
\caption{Trapping horizons for  $\varepsilon=1$, $\omega=g=1$ (left) and for  $\varepsilon=-1$, $\omega=g=1$ (right).
We have set $k=1$ and $r_0=0$. 
The dotted lines for the left figure represent singularities $u=u_{\rm s}^{(\pm )}$. 
The curve  $u=u_{\rm TH}^{(+)}$ is immaterial in the present discussion, since it lies outside the 
 domain of Lorentz signature.
For $\omega<1/2$, the allowed region $R^2>0$ is entirely untrapped.  
The black dot at $u=r=0$ for $\varepsilon=-1$ denotes the instantaneous singularity.
}
\label{fig:TH}
\end{center}
\end{figure}

For the phantom case $\varepsilon=-1$, the trapped region is bounded 
$-\sqrt{(\sqrt{1+4\omega^2}-1)/2}\le gr\le \sqrt{(\sqrt{1+4\omega^2}-1)/2}$. 
The trapping horizon and the throat intersect at the singular point $r=0$ with $u\to 0$.
A typical behavior of trapping horizon in $u-r$ diagram is shown in figure \ref{fig:TH}.

\subsection{Causal structure}

\begin{figure}[t]
\begin{center}
\includegraphics[width=14cm]{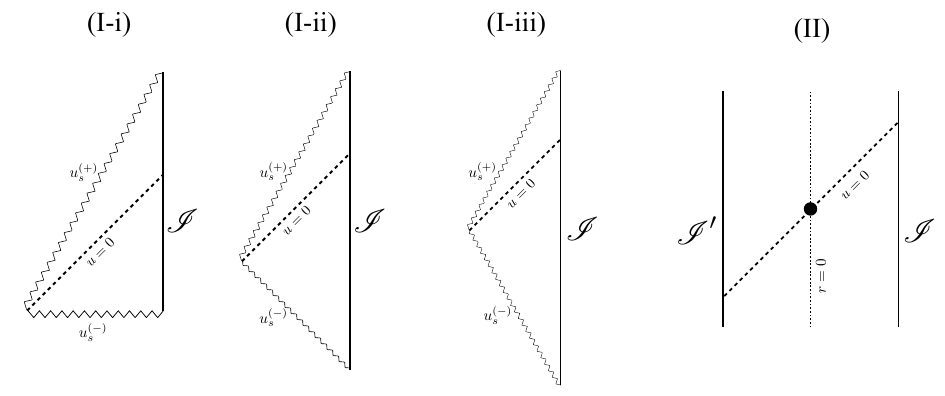}
\caption{Global causal structures of the Petrov D Robinson-Trautman solution (\ref{petrovDsym}) for 
(I) $\varepsilon=+1$ and (II) $\varepsilon=-1$ with $k=1$ and $\omega>0$. The case for a normal scalar field 
splits further into (I-i) $\omega>1/2$, (I-ii) $\omega =1/2$ and (I-iii) $\omega<1/2$.  
 Zigzag lines in (I) and the black dot in (II) denote  curvature singularities. }
\label{fig:PD}
\end{center}
\end{figure}

We are now ready to elucidate the global causal structure 
of the spherical Robinson-Trautman solution (\ref{petrovDsym}).  
For finite $u$, the spacetime is obviously asymptotically AdS. 
Thus the infinity $\mas I$ is timelike. For $\varepsilon=+1$, 
the singularity $u=u_{\rm s}^{(+)}$ is timelike, whilst the  singularity $u=u_{\rm s}^{(-)}$
is timelike for $\omega<1/2$, null for $\omega=1/2$ and spacelike for $\omega>1/2$. 
We visualize the corresponding Penrose diagrams  in (I-i)--(I-iii) in figure~\ref{fig:PD}.

In the phantom case, the spacetime is regular except for the 
instantaneous singularity at $r=0$ with $u\to 0$. The spacetime is asymptotically AdS for $r\to \pm \infty$. 
The Penrose diagram is illustrated in (II) in figure~\ref{fig:PD}. 
We have also performed a numerical study of radial null geodesics to 
substantiate the structure of another asymptotic infinity $\mas I'$ for $\varepsilon=-1$.
Apart from the singular point,  the spacetime describes a dynamical wormhole, connecting two
detached AdS extrema $\mas I$ ($\phi=0$) and $\mas I'$ ($\phi=\phi_{+1}$).

\subsection{Linearized solution of equation (\ref{Keq})}

In the above, we have spelled out the global behavior of the spherical solution (\ref{petrovDsym}). 
It is instructive to see the behavior of the solution to (\ref{Keq}) away from the static and spherical symmetry. 
To this end, we set $P=(1+z\bar z/2)\mathfrak f$ and focus on the behavior of $\mathfrak f$. 
We can verify  
\begin{align}
\label{}
K=\mathfrak f^2 \left(1+\Delta_{S^2}\ln \mathfrak f\right)\,, \qquad 
\Delta_2=\mathfrak f^2 \Delta_{S^2}\,, 
\end{align}
where $\Delta_{S^2}$ is the Laplace operator of $S^2$. 
Working in the gauge $U_1(u)=1$ and $U_0(u)=r_0^2$ ($r_0>0$), 
equation (\ref{Keq}) is  brought into 
\begin{align}
\label{}
\mathfrak f^2 \Delta_{S^2}\left[\mathfrak f^2 \left(1+\Delta_{S^2}\ln \mathfrak f\right)\right] 
+2\varepsilon r_0^2 \mathfrak f^2\partial_u^2(\mathfrak f^{-2})=0\,.
\end{align}
Upon linearization
$\mathfrak f=1+\mathfrak f_{(1)}$, $\mathfrak f_{(1)}$ obeys
\begin{align}
\label{}
\Delta_{S^2}\left(2\mathfrak f_{(1)}+\Delta_{S^2}\mathfrak f_{(1)}\right)-4\varepsilon r_0^2\partial_u^2
\mathfrak f_{(1)}=0\,. 
\end{align}
Expanding in terms of the spherical harmonic as $\mathfrak f_{(1)}=\mathfrak f_{(1)lm}(u)Y_{lm}$, 
we find
\begin{align}
\label{instability}
\mathfrak f_{(1)lm} = \exp\left(\pm \frac{\sqrt{\varepsilon l(l-1)(l+1)(l+2)}}{2r_0}u\right)\,,
\end{align}
for $l(l-1)\ne 0$, and 
$\mathfrak f_{(1)lm}=\omega u+u_0$ for $l(l-1)=0$. 
The latter replicates the $u$-dependence of (\ref{petrovDsym}) for the spherical mode. 
Contrary to the $\Lambda$-vacuum case, we have two independent modes, since 
(\ref{Keq}) involves the second $u$-derivative. 
For $\varepsilon=+1$, one mode is  damping, while the other mode is inevitably growing. 
This implies that  the static solution (\ref{PetrovDstatic}) suffers from instability. 
For $\varepsilon=-1$, on the other hand, $\mathfrak f_{(1)lm}$ displays the sinusoidal behavior. 
Although the growing behavior does not develop, 
this also signals the instability of the static wormhole solution, since the slight deviation from the 
background solution fails to decay at asymptotically late times to a stationary solution.

\section{Concluding remarks}
\label{sec:summary}

The Robinson-Trautman family of non-vacuum solutions have been attracted much attention both
from physical and mathematical perspectives. 
The algebraic classification of non-vacuum solutions has been achieved in \cite{Podolsky:2016sff}, on which 
our paper is based. 
Our present work constitutes the first classification of spacetimes admitting a null geodesic congruence which 
is shear-free and twist-free in Einstein's gravity with a scalar field, 
under the additional assumptions $\Psi_1=\Phi_{01}=\Phi_{02}$.
This generalizes the upshots in \cite{Tahamtan:2015sra,Tahamtan:2016fur} substantially. 
In the case of massless scalar field, this class of solutions falls into three distinct families. 
The class (I-a) corresponds to a simple generalization of the Robinson-Trautman family with a scalar hair (\ref{sol0}) obeying a master equation (\ref{Keq0}). 
A notable discovery that we have made in this paper is the revelation that  the Robinson-Trautman equation (\ref{Keq0cov}) for class (I-a) follows from the first-order Ricci flow equation \cite{Hamilton,Perelman:2006un} as (\ref{RicciFlowderv}). This bears a resemblance to 
BPS solutions which preserve supersymmetry, where the integrability of the first-order Killing spinor equation yields 
the second-order Einstein's equations \cite{Maeda:2011sh}. An analogous phenomenon occurs for the  domain wall  solutions~\cite{Townsend:2007nm}.  
The underlying fundamental reasons behind these observations warrant further investigation.

The class (I-b) stands for a non-asymptotically flat solution (\ref{Ibsol}) characterized by two functions (\ref{eqIb}). 
The classification of Petrov-D, along with the generalization to include the runaway potential, can be found in appendix~\ref{sec:classIb}. 
We have demonstrated that the governing equations for class (I-b) conform exactly to Perelman's pair of Ricci flow equations. As far as we know, our result provides the first example for exact solutions to Einstein's equations admitting (a pair of) the Ricci flows. Up to now, Ricci flow equations have shown up in some different contexts of gravity research. 
For instance,  in \cite{Headrick:2006ti}, the saddle point of the Euclidean path integrals was studied using the Ricci flow.  
 The Ricci flow equation also appeared in the  analysis of topology-changing transition between black strings and localized black holes 
in the large $D$ expansion of general relativity~\cite{Emparan:2019obu}. 
The emergence of the flow equations is clearly advantageous in exploring the global structure of the solutions \cite{Lukacs:1983hr}.
We anticipate that 
our analysis sheds a new light on the relation between the gradient flow equations and Einstein's equations. 

The class (II) represents 
a dynamical solution (\ref{classII}) possessing ${\rm SO}(3)$, ${\rm ISO}(2)$ or ${\rm SO}(1,2)$ symmetry. 
Unfortunately, the field equations for class (II)  reaches a level of considerable complexity which 
hinders our ability to obtain the exact solution. Regardless of this hurdle, we were able to construct a new solution 
via the AdS/Ricci-flat correspondence, by exploiting the AdS-Roberts solution as a seed solution. 
The derived solution for $\varepsilon=-1$ may be identified as a dynamical wormhole. 
The detail is relegated to appendix~\ref{sec:classII}.

Building upon the asymptotically flat class (I-a) solution, we constructed the corresponding asymptotically AdS Robinson-Trautman solution. 
We performed an exhaustive classification of the Petrov-D solutions within this class based on the pioneering work \cite{Podolsky:2016sff}, 
which has accomplished an important job of linking the Petrov classification to the matter content of the spacetime. The first family corresponds to the C-metric, 
as anticipated from the $\Lambda$-vacuum case~\cite{Stephani:2003tm}. For $\varepsilon=+1$, the C-metric describes a naked singularity~\cite{NT1}, 
whilst the C-metric for $\varepsilon=-1$ represents accelerated wormholes in AdS~\cite{NT2}. The second family  
(\ref{petrovDsym}) is dynamical with ${\rm SO}(3)$, ${\rm ISO}(2)$ or ${\rm SO}(1,2)$ symmetry. We  thoroughly examined the causal structure of the 
second family of Petrov-D solutions. Our outcome is visually delineated as the Penrose diagrams in figure~\ref{fig:PD}. 
We found that the phantom solution is not entirely regular but admits an instantaneous spacetime singularity. 
The presence of an instantaneous singularity in a highly symmetrized spacetime might not be generic and could potentially vanish in certain physical scenarios.

It is tempting to expect that the Einstein-scalar gravity may harbor a pretty rich spectrum of spacetimes yet to be discovered. 
Clearly, there exist solutions to (\ref{Keq}) which cannot be expressed as a Ricci flow. It is interesting to seek explicit solutions of this form.
It is also desirable to generalize this result to include the Maxwell field, 
which is expected  to alleviate the instability (\ref{instability}).
A further exploration of corresponding solutions in dimensions 
$D>4$ is of great interest. Work along this direction holds significant promise.

\subsection*{Acknowledgements}
MN would like to thank Ryotaku Suzuki for drawing some references to our attention and for useful comments on Ricci flow.
We appreciate anonymous referees for a critical reading of the manuscript and a number of suggestions which have helped us to improve the paper.
The work of MN is partially supported by MEXT KAKENHI Grant-in-Aid for Transformative Research Areas (A) through the ``Extreme Universe'' collaboration 21H05189 and JSPS Grant-Aid for Scientific Research (20K03929). 
The work of TT is supported by JSPS KAKENHI Grant-Aid for Scientific Research (JP18K03630, JP19H01901) and for Exploratory Research (JP19K21621, JP22K18604).

\appendix
\renewcommand{\theequation}{A.\arabic{equation}}
\setcounter{equation}{0}

\section{Ricci flow}
\label{sec:RF}

This appendix collects some technical details of the Ricci flow. 
Let $g_{ij}$ be a  Riemannian metric 
on an $n$-dimensional compact manifold $M$ without boundary. 
Hamilton incepted the following first-order evolution equation of the metric \cite{Hamilton}
\begin{align}
\label{app:RF}
\frac{\partial }{\partial t}g_{ij}=&\, -2 R_{ij} \,,
\end{align}
where $t$ is the auxiliary flow parameter. 
This is a nonlinear, parabolic partial differential equation. 
Along the Ricci flow, we have 
\begin{align}
\label{}
\dot R_{ij}
=& \, \Delta R_{ij}-2 R^k{}_{i}R_{jk}+2 R_{ikjl}R^{kl} \,, \\
\dot R=& \, \Delta R +2 R_{ij}R^{ij} \,,
\label{dotR}
\end{align}
where $\Delta=\nabla_i\nabla^i$ is the Laplace operator and 
the dot denotes the partial derivative with respect to $t$. 
The Ricci flow  has been originally  studied with the aim of 
proving the geometrization conjecture for closed three-manifolds. 
The methodology of the Ricci flow has also been discussed in the context of  quantum field theory, serving as an approximation of 
renormalization group flow for the two-dimensional nonlinear sigma model. 
Readers interested in a physicist-oriented review may refer to \cite{Woolgar:2007vz}.

Perelman introduced another evolution equation, which, 
after DeTurck's trick, takes the form  \cite{Perelman:2006un}
\begin{align}
\label{feq}
\frac{\partial }{\partial t} f=&\, -\Delta f+(\nabla f)^2 -R \,,
\end{align}
where $f$ is a scalar function on $M$. 
In terms of $f$, 
Perelman defined the following ``entropy functional'' \cite{Perelman:2006un}
\begin{align}
\label{calF}
\ma F= \int_M \left(R+(\nabla f)^2\right)e^{-f}\D V \,, 
\end{align}
where $\D V=\sqrt{{\rm det}g} \D ^n x$ is the volume element. This is the string-frame 
NSNS supergravity action, if we identity $f$ as  (twice of) the dilaton field \cite{Polchinski:1998rq}. 
The entropy functional evolves along the Ricci flow as 
\begin{align}
\label{}
\frac{\D \ma F}{\D t} = \int _M\left[\dot R+2 \nabla^i f\nabla_i \dot f+\dot g^{ij}\nabla_i f\nabla_j f
-\dot f (R+(\nabla f)^2)-\frac 12 g_{ij}\dot g^{ij} (R+(\nabla f)^2)\right]e^{-f} \D V\,.
\end{align}
Using (\ref{dotR}) and the identity
$\nabla^i f\nabla_i \dot f e^{-f}=\nabla_i(\nabla^i fe^{-f}\dot f)-(\Delta f- (\nabla f)^2)e^{-f}\dot f$, 
we get 
\begin{align}
\label{}
\frac{\D \ma F}{\D t} = &\int _Me^{-f } \Big[2 R_{ij}(R^{ij}+\nabla^i f\nabla^j f)+2(\Delta f)^2+2 R(\Delta f-(\nabla f)^2)+(\nabla f)^2 ((\nabla f)^2-3\Delta f)\Big]\D V \,,
\end{align}
where we have dropped the total divergence and 
$e^{-f}\Delta R=\nabla_i (e^{-f}\nabla^i R+e^{-f}\nabla^i f R)-R e^{-f}(\Delta f-(\nabla f)^2)$
has been used. Availing ourselves of the following identity 
\begin{align}
\label{}
e^{-f}|R_{ij}+\nabla_i \nabla_j f|^2 =&\,  e^{-f}\left[R_{ij}(R^{ij}+\nabla^i f\nabla^j f)+(\Delta f)^2
+R(\Delta f-(\nabla f)^2)+\frac 12 (\nabla f)^2 ((\nabla f)^2-3 \Delta f)
\right]\notag \\
&
+\nabla^i \left[e^{-f }\left(2  R_{ij}\nabla^j f+ \nabla^j f \nabla_i\nabla_j f-\nabla_i f \Delta f+\frac 12 (\nabla f)^2 \nabla_i f
- \nabla_i f R \right)\right]\,,
\end{align}
it turns out that the entropy functional is non-decreasing along the Ricci flow
\begin{align}
\label{dotF}
\frac{\D \ma F}{\D t } = \int_M 2 e^{-f}|R_{ij}+\nabla_i \nabla_j f|^2 \D V \ge 0 \,. 
\end{align}

Defining $\varrho \equiv e^{-f}$, the evolution equation (\ref{feq}) of $f$ is rewritten into the 
time-reversed heat diffusion equation 
\begin{align}
\label{}
\frac{\partial \varrho}{\partial t}=-\Delta \varrho+R\varrho\,.
\end{align}
It follows that the ``total charge''  $\int_M \varrho \D V=\int_M e^{-f} \D V$
is conserved 
\begin{align}
\label{}
\frac{\D }{\D t}\int \varrho \D V
=\int _M\left(\dot \varrho+\frac 12 g^{ij}\dot g_{ij}\varrho\right)\D V
=-\int_M \Delta \varrho \D V =0 \,.
\end{align}
One can thus normalize $\int_Me^{-f}\D V=1$. 
Using the Cauchy-Schwarz inequality $|R_{ij}+\nabla_i \nabla_j f|^2\ge \frac 1n(R+\Delta f)^2$, 
we can sharpen the lower bound of (\ref{dotF}) as 
\begin{align}
\label{}
\frac{\D \ma F}{\D t } \ge \frac 2n \int_M e^{-f}(R+\Delta f)^2 \D V
\ge  \frac 2n \left(\int_Me^{-f}(R+\Delta f) \D V\right)^2 =\frac 2n \ma F^2 \,. 
\end{align}
At the second inequality, Schwarz's inequality 
$\left(\int_M e^{-f}(R+\Delta f)^2 \D V\right)\cdot \left(\int _Me^{-f}\D V\right)\ge \left(\int _Me^{-f}(R+\Delta f) \D V\right)^2$
and the constraint $\int_Me^{-f}\D V=1$ have been used.

One can also define the following Boltzmann entropy \cite{reviewRF}
\begin{align}
\label{calN}
\ma N \equiv -\int_M \varrho \ln  \varrho \D V \,.
\end{align}
Along the Ricci flow, we have 
\begin{align}
\label{}
\frac{\D \ma N }{\D t} = \int_M \left[\dot f \varrho (\ln \varrho+1)+R \varrho\ln \varrho\right]\D V
=-\int_M e^{-f}\left(R+(\nabla f)^2\right)\D V=-\ma F \,. 
\end{align}
These functionals have played a central role in proving Thurston's conjecture \cite{Perelman:2006un}.

In the case of $n=2$, the integration of the scalar curvature gives the 
Euler characteristic of $M$ 
\begin{align}
\label{}
\chi =\frac 1{4\pi} \int _M R \D V \,, 
\end{align}
which is a topological invariant. A direct computation shows
\begin{align}
\label{chidot}
\frac{\D \chi}{\D t}= \frac 1{4\pi} \int _M \left(\dot R+\frac 12 g^{ij}\dot g_{ij}R\right)\D V
=\frac 1{4\pi} \int _M \Delta R \D V=0\,. 
\end{align}
It turns out that the topology of the space is unchanged along the Ricci flow. This is the property common to
the Calabi flow~\cite{Lukacs:1983hr}.

In the following, we shall enumerate some exact solutions to Ricci flows
(\ref{app:RF}) and (\ref{feq}) in two dimensions. These solutions are 
promoted to class (I-a) solutions (\ref{sol0}), (\ref{RTmetric2}) with $\varepsilon=+1$, and (I-b) solutions (\ref{Ibsol}), (\ref{Ibsolpot}) in a 
straightforward fashion.

\subsection{Configuration of constant entropy functional}

The class (I-b) solution (\ref{Ibsol2}), which will be derived in appendix~\ref{sec:classIb}, can be identified as a dynamical black hole, despite 
the unfamiliar asymptotic structure. Since the non-decreasing property (\ref{dotF}) of the entropy functional is 
reminiscent of the area-increasing theorem of event horizon~\cite{Hawking:1973uf}, 
one might be inclined to expect  that the Bekenstein-Hawking entropy of this solution 
would be related to the entropy functional (\ref{calF}). 

To clarify this, 
let us discuss the configuration in which the entropy functional is constant along the Ricci flow $\D \ma F/\D t=0$. 
From (\ref{dotF}), this occurs for 
\begin{align}
\label{}
R_{ij}+\nabla_i \nabla_j f=0\,. 
\end{align}
This equation exactly agrees with Einstein's equation derived from the action (\ref{calF}). 
Focusing on two dimensions and writing $g_{ij}^{(2)}\D x^i \D x^j=2 P^{-2}\D z \D \bar z$, 
$z\bar z$-component of this equation amounts to $P=e^{-f/2}$, where additional holomorphic 
and anti-holomorphic functions are absorbed by $z\to \ti z(z)$. Then, $zz$-component of 
the above equation is combined with (\ref{app:RF}) and (\ref{feq}) to give
\begin{align}
\label{}
f=-\log \left(b_1 z+\bar b_1 \bar z-2 |b_1|^2 t+b_2 \right)\,,
\end{align}
where $b_1$ is complex and $b_2$ is real. 
It turns out that the four-dimensional spacetimes corresponding to constant entropy functional are  time-dependent. 
Then, this functional $\ma F$ is irrelevant to the Bekenstein-Hawking entropy, since the latter 
should be constant for stationary state.

\subsection{Logarithmic diffusion equation}

In the body of text, we have encountered a two-dimensional Ricci flow equation, which has been 
extensively studied hitherto in the situation where an isometry is present. 
Writing $g^{(2)}_{ij}\D x^i \D x^j =2 P^{-2}\D z \D \bar z$, 
we assume $P=P(t, z, \bar z)$  is independent of ${\rm Im}(z)$. Then, 
$V(t, (z+\bar z)/\sqrt 2)=P^{-2}$ satisfies the logarithmic diffusion equation
\begin{align}
\label{}
\frac{\partial }{\partial t}V=  \frac{\partial^2 }{\partial x^2}\ln V \,, 
\end{align}
where $x\equiv (z+\bar z)/\sqrt 2$. 
The logarithmic diffusion equation emerges when studying the
thermalized electronic cloud. Assuming $f=f(t ,x )$, $f$ obeys
\begin{align}
\label{}
\frac{\partial f}{\partial t}=\frac 1{V} \left[\frac{\partial^2 }{\partial x^2}\ln V+e^{f}\frac{\partial^2 }{\partial x^2}(e^{-f})\right]\,.
\end{align}
The Gauss curvature is given by
\begin{align}
\label{}
K=-\frac 1{2V}\frac{\partial^2}{\partial x^2 } \ln V =-\frac 12 \frac{\partial}{\partial t}\ln V \,. 
\end{align}
We shall now present some exact solutions for the two-dimensional Ricci flow.

\subsubsection*{Separable solution}

Assuming the separable form $V(t, x)=V_{1}(t)V_2(x)$, we arrive at 
\begin{align}
\label{}
 V(t, x)= 2 a_0 t\sec^2\left(\sqrt{a_0}x\right)\,,
\end{align}
for which $K=-1/(2t)$, which asymptotically tends to vanish as $t\to \infty$. 
This implies that the solution approaches infinitely close to the plane symmetric one. Nevertheless, it should be remarked that 
this does not give rise to the topology change of the surface, as inferred from  (\ref{chidot}).

Assuming $f(t,x)=f_1(t)+f_2(x)$, these functions are 
obtained by quadrature. Since the expression is pretty messy, 
we do not write down the results here. 

\subsubsection*{de Gennes solution}

The de Gennes solution describing the spreading of microscopic droplets~\cite{deGennes} is given by
\begin{align}
\label{}
 V(t, x)=\frac{1}{1+a_1 e^{a_2 (x+a_2 t)}} \,, 
\end{align}
where $a_1$ and $a_2$ are constants. The Gauss curvature is $K=a_1a_2^2e^{a_2(a_2t+x)}/[2(1+a_1 e^{a_2(a_2 t+x)})]$, leading to 
$K\to a_2^2/2>0$ as $t\to \infty$.
Assuming $f$ depends only on a combination $x+a_2 t$, we find 
\begin{align}
\label{}
f(t, x)=a_2 (x+a_2 t)+a_3 -\ln \left\{(1+a_1e^{a_2 (x+a_2 t)})\left[a_1a_2+a_4 \ln \big(1+a_1e^{a_2 (x+a_2 t)}\big)\right]\right\}\,.
\end{align}

\subsubsection*{King-Rosenau solution}

The King-Rosenau solution reads \cite{King,Rosenau}
\begin{align}
\label{}
V(t ,x)=\frac 1{\omega}\left[\tanh (x-\omega t)-\tanh(x+\omega t)\right]\,, 
\end{align}
where $\omega$ is a constant. 
The Gauss curvature reads $K=-\omega [1+\cosh(2x)\cosh(2\omega t)]/[\sinh(2\omega t)(\cosh(2x)+\cosh(2\omega t))]$, 
which approaches to zero as $t\to \infty$.
It seems that there is no solution of the form 
$f(t, x)=f_0 (x-\omega t)+f_1 (x+\omega t)$.

\section{Class I-b solution}
\label{sec:classIb}

The class (I-b) solution derived in the body of text is given by (\ref{Ibsol}). 
Since this class of solutions appears to be new, we are now delving into this  in more detail. 
A straightforward calculation shows that the curvature tensors satisfy
\begin{align}
\label{RrelIb} 
R_{\mu\nu}R^{\mu\nu}=R^2 \,, \qquad 
R_{\mu\nu\rho\sigma}R^{\mu\nu\rho\sigma}=3 R^2 \,,
\end{align}
where the scalar curvature is given by $R=\left(H_1-2r \partial_u \ln P\right)/(2r^2)$.
It turns out that the central point $r=0$ is a curvature singularity. 
Since $R$, $R_{\mu\nu}R^{\mu\nu}$ and $R_{\mu\nu\rho\sigma}R^{\mu\nu\rho\sigma}$ are entirely 
distinct spacetime invariants, there are no a priori relationships between these quantities as (\ref{RrelIb}). 
In some special situations such as a homogeneous G\"odel spacetime~\cite{Hawking:1973uf}, these relations may hold true because curvature invariants must be constant
and the solution is characterized by a single parameter. 
In the present case, however, the above relation (\ref{RrelIb}) 
is quite unexpected, since the solution is neither specified by a 
single parameter nor $R$ is constant. 
Although the aforementioned relationships might suggest the presence of specific differential or algebraic symmetries, 
the fundamental reason behind their validity is not currently clear to us.

Taking the null tetrad as
\begin{align}
\label{}
l =\frac{\partial}{\partial r}\,, \qquad 
n=\frac{\partial}{\partial u}-\frac 12 \left(
H_1-2r \partial_u \ln P \right)\frac{\partial}{\partial r}\,, \qquad 
m=\frac{P}{\sqrt{\ell r}} \frac{\partial}{\partial z}\,, 
\end{align}
the nonvanishing Weyl scalars for the class (I-b) solution (\ref{Ibsol}) read
\begin{align}
\label{}
\Psi_2=& \,\frac{\ell H_1 -2 r K}{12\ell r^2}\,, \\
\Psi_3=& \,-\frac{P}{4\sqrt{\ell} r}\left(\frac{L}{P^2}+\frac{\partial_{\bar z}H_1}{r}\right)\,, \\
\Psi_4=& \, \frac{2P\partial_{\bar z}H_1 \partial_{\bar z} P+P^2 \partial_{\bar z}^2 H_1-r \partial_{\bar z} L}{2\ell r}\,, 
\end{align}
where we have denoted $K\equiv 2P^2 \partial_z\partial_{\bar z}\ln P$ and $L\equiv 2P^2 \partial_{\bar z}\partial_u \ln P$
as before. 

\subsection{Petrov D solutions}

It is instructive to undertake the exhaustive classification of the Petrov D for the class (I-b) solution. 
Plugging the above equations of Weyl scalars into (\ref{PetrovD}), the Petrov D requirement amounts to three conditions 
$D_{(0)}=D_{(1)}=D_{(2)}=0$, where 
\begin{align}
\label{IbD0}
D_{(0)}=&\, -2H_1 \partial_{\bar z}H_1  \partial_{\bar z}P+P[(\partial_{\bar z}H_1)^2-H_1\partial_{\bar z}^2 H_1] \,, \\
\label{IbD1}
D_{(1)}=&\, 2\ell L\partial_{\bar z}H_1+\ell H_1 \partial_{\bar z}L+2 PK (2\partial_{\bar z}H_1\partial_{\bar z}P+P\partial_{\bar z}^2 H_1)\,, \\
\label{IbD2}
D_{(2)}=&\, \ell L^2 -2 P^2 K \partial_{\bar z}L \,.
\end{align}
The classification of the solutions proceeds according to $\partial_{\bar z}H_1 = 0$ or not.

\subsubsection{$\partial_{\bar z}H_1 \ne 0$ case}

For $\partial_{\bar z}H_1 \ne 0$, 
integration of (\ref{IbD0}) implies 
\begin{align}
\label{H1zb}
  \partial_{\bar z}H_1=\frac{d_0(u,z)}{P^2}H_1 \,,
\end{align}
where $d_0(\ne 0)$ is independent of $\bar z$. 
Other equations yield 
\begin{align}
\label{KL}
L=&\, -\frac{2d_0}{\ell }K \,, \\
\label{KzbIb}
\partial_{\bar z}K=& \, -\frac{d_0}{P^2}K \,. 
\end{align}
The $u$-derivative of $K=2P^2 \partial_z\partial_{\bar z}\ln P$ implies
\begin{align}
\label{KuIb}
\partial_u K =2 K \partial_u \ln P-\frac 2 \ell  P^2 \partial_z \left(\frac{d_0 K}{P^{2}}\right)\,. 
\end{align}
The integrability condition $(\partial_u \partial_{\bar z} -\partial_{\bar z}\partial_u )K=0$ gives 
\begin{align}
\label{PzIb}
 \partial_{z} P=\frac{\partial_z d_0}{2d_0}P+\frac{\ell}{4d_0^2}\partial_u \left(\frac{d_0}{P^{2}}\right)P^3 \,. 
\end{align}
Taking the complex conjugation of (\ref{KzbIb}), the compatibility condition 
$(\partial_z \partial_{\bar z} -\partial_{\bar z}\partial_z )K=0$ yields 
\begin{align}
\label{}
\frac{\partial_u d_0}{d_0}= \frac{\partial_u \bar d_0}{\bar d_0}\,. 
\end{align}
It follows that $d_0$ admits the separable structure 
$d_0(u, z)=d_{00}(u)d_{01}(z)$ with real $d_{00}$.  
Substituting the expression of $K$ in (\ref{KIb}) into (\ref{KuIb}), $d_{00}(u)$ turns out to satisfy
\begin{align}
\label{}
d_{00}d_{00}''-2d_{00}'{}^2=0\,,
\end{align}
where we have employed the gauge $\phi_0=0$.  
This equation is solved by $d_{00}(u)=u_0/u$, where $u_0$ is a constant and another integration constant has been 
eliminated by the shift $u\to u+{\rm constant}$.  
Then, $P=P(u, z, \bar z)$ satisfies 
\begin{align}
\label{Pzeq}
u \partial_u \left(\frac 1{u P^2}\right)+\frac{2u_0 }{\ell u}\partial_z \left(\frac{d_{01}(z)}{P^2}\right)=0 \,. 
\end{align}
Performing  the transformation $z\to \ti z(z)$ with 
$d_{01}(z)\propto \D z/\D \ti z$ [see (\ref{Ibtr})], one can set $d_{01}(z)=1$. 
Then, equation (\ref{Pzeq}) is solved as 
\begin{align}
\label{}
P(u,z,\bar z)=\sqrt{\frac{u_0}{u}}P_0(s) \,, \qquad s\equiv \ln \left(\frac{u}{u_0}\right)-\frac{\ell}{2u_0}(z+\bar z) \,.
\end{align}
The Ricci flow equation 
(\ref{KIb}) is recast into 
\begin{align}
\label{}
\frac{\D ^2}{\D \eta^2} \ma P (\eta) -\frac{2u_0}{\ell }\frac{\D }{\D \eta} \ma P(\eta)-\frac{2u_0}{\ell}=0 \,, 
\end{align}
where $\D \eta/\D s=P_0^{-2}$ and $\ma P=P_0^{-2}$ have been introduced. 
This equation is solved as
\begin{align}
\label{}
\ma P(\eta ) =-\eta +b_1e^{2u_0\eta/\ell}+b_2 \,,
\end{align}
where $b_1$ and $b_2$ are constants. 
Then, the solution to (\ref{H1zb}) and (\ref{H1eq}) is 
\begin{align}
\label{}
H_1 =\frac{b_3u_0}{u} e^{-2u_0\eta /\ell} \,,
\end{align}
where $b_3$ is a constant. Writing $z={\rm Re}(z)+i (u_0/\ell) \varphi$, 
it turns out that the metric and the scalar field are given by
\begin{subequations}
\label{IbDsol1}
\begin{align}
\D s^2 =&\,-2 \D u \D r-\frac{4u_0 r }{\ell}\D u \D \eta +\frac{u_0}{u}\left(-b_3e^{-2u_0\eta /\ell}+\frac{2(b_2-\eta)}{\ell}r\right)\D u^2
\notag \\
&\, +2 \frac{u_0}{\ell }u r \left(\frac{\D \eta^2}{\ma P(\eta)}+\ma P(\eta)\D \varphi^2\right)\,, \\
\phi=&\, \frac 1{\sqrt 2}\ln \left(\frac r\ell\right)\,.
\end{align}
\end{subequations}
By the direct integration of the Killing equation, it can be shown that the above solution (\ref{IbDsol1}) 
allows only a single Killing vector $\partial/\partial \varphi$. This kind of solutions exists 
in the Einstein-Maxwell case [see (28.46) of \cite{Stephani:2003tm}], but does not arise in the vacuum case 
[Section 26.2 of \cite{Stephani:2003tm}]. 


\subsubsection{$\partial_{\bar z}H_1 =0$ case}

When $\partial_{\bar z}H_1 =0$, we have $H_1=H_1(u)$. 
For $H_1\ne 0$, the vanishing of 
(\ref{IbD1}) and (\ref{IbD2}) implies 
$L=0$, leading to $P=P_0(u)P_1(z, \bar z)$. 
Insertion of this into the first of (\ref{eqIb}), 
we get 
\begin{align}
\label{}
k P_0^3 = \ell \partial_u P_0 \,, \qquad 2 P_1^2 \partial_z \partial_{\bar z} P_1 =k \,,  
\end{align}
where $k$ is the separation constant normalized as $k=0, \pm 1$. 
We thus find that $2 P_1^{-2}\D z \D \bar z$ is the space of constant Gauss curvature $k$ 
and 
\begin{align}
\label{}
 P_0(u) = \sqrt{\frac{1}{2(b_1-k u/\ell )}}\,. 
\end{align}
The second equation of (\ref{eqIb}) is solved as 
\begin{align}
\label{}
H_1(u)=\frac{b_2}{b_1-k u/\ell} \,. 
\end{align}
Thus, the solution reads
\begin{align}
\label{Ibsol2}
\D s^2=-2 \D u \D r -\frac{b_2-k r/\ell }{b_1-k u/\ell} \D u^2+2 \ell r (b_1-k u/\ell ) \D \Sigma_k^2 \,, \qquad 
\phi=\frac 1{\sqrt 2}\ln \left(\frac{r}{\ell}\right)\,.
\end{align}
This solution has a curvature singularity at $r(b_1-k u/\ell)=0$ where the areal radius vanishes. 
It can be also checked that 
\begin{align}
\label{}
\nabla_{(\mu} \zeta_{\nu)}= k g_{\mu\nu} \,, \qquad 
\zeta \equiv  -2\ell \left(b_1-\frac{ku}{\ell}\right)\frac{\partial}{\partial u}\,.
\end{align}
It follows that $\zeta^\mu$ is a Killing vector for $k=0$ and a homothetic Killing vector for $k\ne 0$. 
In the former case, the solution is static, while in the latter case, the solution is self-similar and conformally static. 
Introducing $U \equiv -(\ell/k )\ln (b_1-ku/\ell)$ for $k\ne 0$,  the metric can be brought into the form 
\begin{align}
\label{}
\D s^2=e^{-k U/\ell} \left[-2 \D U \D r -\left(b_2-\frac{kr}{\ell}\right)\D U^2+2\ell r \D \Sigma_k^2 \right]\,,
\end{align}
which satisfies $g_{\mu\nu}\to e^{-kC/\ell}g_{\mu\nu}$ for $U\to U+C$. 
Although the causal structure can be pursued in an analytic fashion, we shall not attempt to do this
here, since this requires as much discussion as section~\ref{sec:D}.

For $H_1=0$, $D_{(0)}=D_{(1)}=0$ is fulfilled automatically. 
The Ricci flow (\ref{eqIb}) gives
\begin{align}
\label{IbPu}
\partial_u P= \frac 1\ell P K \,,
\end{align}
while 
 $D_{(2)}=0$ yields 
\begin{align}
\label{}
P[(\partial_{\bar z} K)^2-K\partial_{\bar z}^2 K]-2 K\partial_{\bar z}K \partial_{\bar z} P=0 \,. 
\end{align}
For $K=0$, this solution is satisfied automatically and the solution reduces to the one (\ref{Ibsol2}) with $k=0$. 
When $K\ne 0$, the above equation is integrated once as 
\begin{align}
\label{IbKzb}
 \partial_{\bar z}K= -\frac{d_0(u, z)}{P^2}K \,,
\end{align}
where $d_0(\ne 0)$ is independent of $\bar z$. 
It follows from the definition $K=2P^2\partial_z\partial_{\bar z}\ln P$  and (\ref{IbPu}) that 
\begin{align}
\label{IbPu2}
\partial_u K=
2K \partial_u \ln P-\frac 2\ell P^2 \partial_z\left(\frac{d_0 K}{P^{2}}\right) \,. 
\end{align}
Since the governing equations (\ref{IbKzb}) and (\ref{IbPu2}) are 
identical to the previous case (\ref{KzbIb}) and (\ref{KuIb}), 
it turns out that the final expression of the metric is (\ref{IbDsol1}) with $b_3=0$, although $b_3\ne 0$ has been 
assumed therein.

\subsection{Runaway potential}

The prescription given in  section \ref{sec:ansatz}, which provides a 
solution in a theory with the potential (\ref{pot:superpotential}), 
does not work for the  class (I-b) family. 
However, an analogous procedure generates a new solution for the 
Einstein-scalar theory (\ref{Lag}) with a runaway potential 
\begin{align}
\label{runaway}
\ma V(\phi)=-\frac{g^2}{2}e^{-\sqrt 2 \phi} \,.
\end{align}
This theory is also obtained by the ${\cal N}=2$ gauged supergravity 
with the prepotential $F=-iX^0X^1$  (\ref{FIV}), provided that one of the 
Fayet-Iliopoulous gauge coupling constants is set to vanish ($g_0=0$). 

The solution reads
\begin{align}
\label{Ibsolpot}
\D s^2=-2 \D r \D u - \left(H_1-2r \partial_u \ln P +g^2 \ell r\right)\D u^2 +\frac{2\ell r }{P^2 }\D z \D \bar z\,, 
\qquad \phi=\frac 1{\sqrt 2} \ln \left(\frac{r}{\ell}\right)\,,
\end{align}
where $P=P(u, z, \bar z)$ and $H_1=H_1(u, z, \bar z)$ obey (\ref{eqIb}). 
Since the scalar field does not vanish for class (I-b), the conformity with the runaway potential is persuasive.
This prescription, instead, does not work for class (I-a) or class (II).

\section{A new class II solution via AdS/Ricci-flat correspondence}
\label{sec:classII}

Let us consider the 
$D$-dimensional gravity with a scalar field
\begin{align}
\label{th1}
\ma L^{(D)}=R{}^{(D)}-\frac 12 \varepsilon (\nabla\phi)^2 \,.
\end{align}
In this theory, 
we center our attention on the spherically symmetric solutions 
\begin{align}
\label{solRF}
\D s_D^2 =\bar g_{ab}(y)\D y^a \D y^b +r(y)^2 \D \Omega_{D-2}^2 \,, \qquad 
\phi=\phi(y) \,,
\end{align}
where $\D \Omega_{D-2}^2$ is the line element of a unit round $(D-2)$-sphere. 
The solution is characterized by the two-dimensional Lagrangian 
\begin{align}
\label{Lag2}
\ma L^{(2)} =r^{D-2}\left[\bar R{}^{(2)}-\frac 12 \varepsilon (\bar \nabla\phi)^2+(D-2)(D-3)\frac{1+(\bar\nabla r)^2}{r^2} \right]\,. 
\end{align}

Let us next consider the 
$d$-dimensional gravity endowed with a massless scalar field and a negative cosmological constant ($\Lambda=-(d-1)(d-2)g^2/2$), 
\begin{align}
\label{th2}
\ti {\ma L}^{(d)}=\ti R{}^{(d)}-\frac 12 \varepsilon (\ti \nabla\phi)^2 +(d-1)(d-2)g^2 \,.
\end{align}
We focus on the warped metric 
\begin{align}
\label{solads}
\D \ti s^2 _d= \ti g_{ab}(y)\D y^a \D y^b +\ti r(y)^2 \D {\bf x}_{d-2}^2\,. 
\end{align}
By the transformation $\ti r=1/(g^2 r)$ with $\ti g_{ab}=(g\ti r)^2 \bar g_{ab}$, 
the solution (\ref{solads}) obeys the field equations derived from the following two-dimensional effective Lagrangian 
\begin{align}
\ti {\ma L}^{(2)}=&\, 
r^{2-d} \left[\bar R{}^{(2)}-\frac 12 \varepsilon (\bar \nabla\phi)^2
+(d-1)(d-2)\frac{1+(\bar \nabla r)^2}{r^2} \right]\,. 
\label{Lag2t}
\end{align}
These Lagrangians (\ref{Lag2}) and (\ref{Lag2t}) are equivalent, modulo the overall constant, 
under the identification~\cite{Caldarelli:2012hy,Caldarelli:2013aaa}
\begin{align}
\label{}
D~~ \leftrightarrow ~~4-d \,.
\end{align}

Using the above recipe, one can generate the solution (\ref{solRF}) in the theory (\ref{th1}) 
out of the seed solution (\ref{solads}) in the theory (\ref{th2}), and vice versa, 
e.g., an AdS wormhole solution from the asymptotically flat one~\cite{Wu:2022gpm}. 

Here, 
we employ the dynamical solution in \cite{Maeda:2015cia}, which is identified as the 
(A)dS generalization of the self-similar Roberts solution \cite{Roberts:1989sk}
\begin{align}
\label{}
\D \ti s_{d}^2 =&\,\frac 1{\Omega(\ti u, \ti v)^2}\left(-2 \D\ti  u \D\ti  v+S_0(\ti u,\ti v) ^2\D \mathbf x _{d-2}^2 \right)\,,\\
\phi=&\, 2\sqrt{\varepsilon \frac{d-3}{d-2}}{\rm arctanh} \left[\sqrt{\frac{\varepsilon C_1}{C_2}}\left(\frac{\ti u}{\ti v}\right)^{\frac{d-2}{2}}\right]\,,
\end{align}
where 
\begin{align}
\label{}
  S_0=\left(C_1 \ti u^{d-2}-\varepsilon C_2\ti v^{d-2}\right)^{\frac 1{d-2}}\,, \qquad 
  \Omega=p_0+p_1\ti u+p_2 \ti v+p_3 \ti u \ti v \,.
\end{align}
Here, the parameters satisfy the constraints 
\begin{align}
\label{}
p_1p_2-p_0p_3=-\frac 12 g^2 \,, \qquad C_1 p_1=C_2p_2 =0 \,. 
\end{align}
Specifically, we set $p_0-1=p_1=p_2=0$ and $p_3=g^2/2$ in what follows.  
Mapping the solution, we obtain a new solution in the Einstein-massless scalar system 
\begin{align}
\label{}
\D s_D^2 =&\, \left(C_1 \ti u^{2-D}-\varepsilon C_2 \ti v^{2-D}\right)^{\frac{2}{D-2}} \left[-2  \D\ti  u \D\ti  v +g^{-2}\left(1+\frac 12 g^2 \ti u \ti v\right)^2\D \Omega_{D-2}^2\right]\,, 
\\
\phi=&\, 2\sqrt{\varepsilon\frac{D-1}{D-2}}{\rm arctanh}\left[\sqrt{\frac{\varepsilon C_1}{C_2}}\left(\frac{\ti u}{\ti v}\right)^{-\frac{D-2}{2}}\right]\,. 
\end{align}
When $D=4$, it turns out that this metric falls into the Class (II) Robinson-Trautman family discussed in section \ref{sec:classIIder} as\footnote{
For $D>4$, the explicit coordinate transformation to (\ref{classII}) is unavailable. 
} 
\begin{align}
\label{IIsol}
\D s_4^2 =&\,-2 g \sqrt{r^2-\varepsilon r_{\rm II}^2 }\D u^2 -2 \D r \D u \notag \\
&
+\frac{\sqrt{r^2-\varepsilon r_{\rm II}^2  }}{2(r+\sqrt{r^2-\varepsilon r_{\rm II}^2  })}\left[
e^{gu}\left(r+\sqrt{r^2-\varepsilon r_{\rm II}^2 }\right)+g^{-1} e^{-g u}
\right]^2\D \Omega_{2}^2\,,
\\
\phi=&\, \sqrt {6\varepsilon}\, {\rm arctanh} \left[\frac{\sqrt{\varepsilon}(r+\sqrt{r^2-\varepsilon r_{\rm II}^2})}{r_{\rm II}}\right]\,,
\end{align}
by
\begin{align}
\label{}
\ti v=\frac{r+\sqrt{r^2-4\varepsilon C_1C_2 g^2 }}{g\sqrt{C_1}}e^{gu} \,, \qquad 
\ti u=2\sqrt{C_1}e^{gu}\,, \qquad 
C_1C_2=\frac{r_{\rm II}^2}{4g^2}\,. 
\end{align}
The Kretschmann invariant for the solution (\ref{IIsol}) is 
\begin{align}
\label{}
R_{\mu\nu\rho\sigma}R^{\mu\nu\rho\sigma}=\frac{15 g^2 r_{\rm II}^4}{(r^2-\varepsilon r_{\rm II}^2)^3}\,.
\end{align}
The solution  (\ref{IIsol}) reduces to the Minkowski spacetime for $r_{\rm II}=0$.
In the case of $\varepsilon=-1$, the solution 
would describe a dynamical wormhole in asymptotically flat spacetime without curvature singularities. 
This solution would be worth investigating further.


\begin{thebibliography}{99}


\bibitem{GS}
J.~N. Goldberg and R.~K.~Sachs, 
Gen. Relat. Grav, {\bf 41}, 2,  433-444, (2009).
doi:10.1007/s10714-008-0722-5


\bibitem{Newman:1961qr}
E.~Newman and R.~Penrose,
J. Math. Phys. \textbf{3}, 566-578 (1962)
doi:10.1063/1.1724257


\bibitem{Kinnersley:1969zza}
W.~Kinnersley,
J. Math. Phys. \textbf{10}, 1195-1203 (1969)
doi:10.1063/1.1664958



\bibitem{RT}
I.~Robinson and A.~Trautman,
Phys. Rev. Lett. \textbf{4}, 431-432 (1960)
doi:10.1103/PhysRevLett.4.431


\bibitem{RT2}
I.~Robinson and A.~Trautman,
Proc. Roy. Soc. Lond. A \textbf{265}, 463-473 (1962)
doi:10.1098/rspa.1962.0036

\bibitem{Tod}
K.~P.~Tod, Class. Quant. Grav. {\bf 6}, 1159 (1989). 
doi:10.1088/0264-9381/6/8/015



\bibitem{FN}
J. Foster, E. T. Newman, 
J. Math. Phys. {\bf 8}, 189-194 (1967). 
doi:10.1063/1.1705185

\bibitem{Lukacs:1983hr}
B.~Luk\'acs, Z.~Perj\'es, J.~Porter and A.~Sebesty\'en,
Gen. Rel. Grav. {\bf 16}, 7, 691-701 (1984). 
doi:10.1007/BF00767861

\bibitem{Vandyck}
M. A. J. Vandyck, 
Class. Quant. Grav. {\bf 2}, 77 (1985). 
doi:10.1088/0264-9381/2/1/008


\bibitem{Vandyck2}
M. A. J. Vandyck, 
Class. Quant. Grav. {\bf 4}, 759 (1987). 
doi:10.1088/0264-9381/4/3/032

\bibitem{Schmidt}
B.~G.~Schmidt, 
Gen. Rel. Grav, {\bf 20},  65 (1988). 
doi:10.1007/BF00759256


\bibitem{Glass}
E. N. Glass, D. C. Robinson, 
J. Math. Phys. {\bf 25}, 3382-3386 (1984).
doi.org/10.1063/1.526107

\bibitem{Hoenselaers}
C. Hoenselaers and W. K. Schief,  
J. Phys. A: Math. Gen. {\bf 25} 601 (1992).
doi:10.1088/0305-4470/25/3/018

\bibitem{Hoenselaers2}
C. Hoenselaers and Z. Perjes, 
Class. Quant. Grav. {\bf 10}, 375 (1993).
doi:10.1088/0264-9381/10/2/019




\bibitem{Rendall}
 A. D. Rendall,  Class. Quant. Grav. {\bf 5}, 1339 (1988).
 doi:10.1088/0264-9381/5/10/012
 
 
 \bibitem{Singleton}
 D. Singleton, Class. Quant. Grav. {\bf 7}, 1333 (1990).
 doi:10.1088/0264-9381/7/8/012


\bibitem{Chrusciel:1991vxx}
P.~Chru\'sciel,
Commun. Math. Phys. \textbf{137}, 289-313 (1991)
doi:10.1007/BF02431882

\bibitem{Chrusciel:1992rv}
P.~T.~Chrusciel,
Proc. Roy. Soc. Lond. A \textbf{436}, 299-316 (1992)
doi:10.1098/rspa.1992.0019

\bibitem{Chrusciel:1992tj}
P.~T.~Chrusciel and D.~B.~Singleton,
Commun. Math. Phys. \textbf{147}, 137-162 (1992)
doi:10.1007/BF02099531




\bibitem{Kozameh:2006hk}
C.~Kozameh, E.~T.~Newman and G.~Silva-Ortigoza,
Class. Quant. Grav. \textbf{23}, 6599-6620 (2006)
doi:10.1088/0264-9381/23/23/002
[arXiv:gr-qc/0607074 [gr-qc]].




\bibitem{Bicak:1995vc}
J.~Bicak and J.~Podolsky,
Phys. Rev. D \textbf{52}, 887-895 (1995)
doi:10.1103/PhysRevD.52.887


\bibitem{Bicak:1997ne}
J.~Bicak and J.~Podolsky,
Phys. Rev. D \textbf{55}, 1985-1993 (1997)
doi:10.1103/PhysRevD.55.1985
[arXiv:gr-qc/9901018 [gr-qc]].




\bibitem{BernardideFreitas:2014eoi}
G.~Bernardi de Freitas and H.~S.~Reall,
JHEP \textbf{06}, 148 (2014)
doi:10.1007/JHEP06(2014)148
[arXiv:1403.3537 [hep-th]].


\bibitem{Bakas:2014kfa}
I.~Bakas and K.~Skenderis,
JHEP \textbf{08}, 056 (2014)
doi:10.1007/JHEP08(2014)056
[arXiv:1404.4824 [hep-th]].


\bibitem{Podolsky:2006du}
J.~Podolsky and M.~Ortaggio,
Class. Quant. Grav. \textbf{23}, 5785-5797 (2006)
doi:10.1088/0264-9381/23/20/002
[arXiv:gr-qc/0605136 [gr-qc]].

\bibitem{Ortaggio:2007hs}
M.~Ortaggio, J.~Podolsky and M.~Zofka,
Class. Quant. Grav. \textbf{25}, 025006 (2008)
doi:10.1088/0264-9381/25/2/025006
[arXiv:0708.4299 [gr-qc]].

\bibitem{Ortaggio:2014gma}
M.~Ortaggio, J.~Podolsk\'y and M.~\v{Z}ofka,
JHEP \textbf{02}, 045 (2015)
doi:10.1007/JHEP02(2015)045
[arXiv:1411.1943 [gr-qc]].


\bibitem{Podolsky:2016sff}
J.~Podolsk\'y and R.~\v{S}varc,
Phys. Rev. D \textbf{94}, no.6, 064043 (2016)
doi:10.1103/PhysRevD.94.064043
[arXiv:1608.07118 [gr-qc]].







\bibitem{Gueven:1996zm}
R.~Gueven and E.~Yoruk,
Phys. Rev. D \textbf{54}, 6413-6423 (1996)
doi:10.1103/PhysRevD.54.6413
[arXiv:hep-th/9609078 [hep-th]].

\bibitem{Tahamtan:2015sra}
T.~Tahamtan and O.~Svitek,
Phys. Rev. D \textbf{91}, no.10, 104032 (2015)
doi:10.1103/PhysRevD.91.104032
[arXiv:1503.09080 [gr-qc]].

\bibitem{Tahamtan:2016fur}
T.~Tahamtan and O.~Svitek,
Phys. Rev. D \textbf{94}, no.6, 064031 (2016)
doi:10.1103/PhysRevD.94.064031
[arXiv:1603.07281 [gr-qc]].









\bibitem{Chandrasekhar:1985kt}
S.~Chandrasekhar,
``The mathematical theory of black holes,''
Oxford University Press, 1985. 


\bibitem{Stewart:1990uf}
J.~M.~Stewart,
``Advanced general relativity,''
Cambridge University Press, 1994,
ISBN 978-0-521-44946-5, 978-0-511-87118-4
doi:10.1017/CBO9780511608179



\bibitem{Stephani:2003tm}
H.~Stephani, D.~Kramer, M.~A.~H.~MacCallum, C.~Hoenselaers and E.~Herlt,
Cambridge Univ. Press, 2003,
ISBN 978-0-521-46702-5, 978-0-511-05917-9
doi:10.1017/CBO9780511535185







\bibitem{Klemm:2011xw}
D.~Klemm,
JHEP \textbf{07}, 019 (2011)
doi:10.1007/JHEP07(2011)019
[arXiv:1103.4699 [hep-th]].


\bibitem{Colleoni:2012jq}
M.~Colleoni and D.~Klemm,
Phys. Rev. D \textbf{85}, 126003 (2012)
doi:10.1103/PhysRevD.85.126003
[arXiv:1203.6179 [hep-th]].


\bibitem{Gnecchi:2013mja}
A.~Gnecchi, K.~Hristov, D.~Klemm, C.~Toldo and O.~Vaughan,
JHEP \textbf{01}, 127 (2014)
doi:10.1007/JHEP01(2014)127
[arXiv:1311.1795 [hep-th]].



\bibitem{NT2}
M.~Nozawa and T.~Torii,
Phys. Rev. D \textbf{108}, no.6, 064036 (2023)
doi:10.1103/PhysRevD.108.064036
[arXiv:2306.15198 [gr-qc]].


\bibitem{Ishibashi:2004wx}
A.~Ishibashi and R.~M.~Wald,
Class. Quant. Grav. \textbf{21}, 2981-3014 (2004)
doi:10.1088/0264-9381/21/12/012
[arXiv:hep-th/0402184 [hep-th]].



\bibitem{Hertog:2004dr}
T.~Hertog and K.~Maeda,
JHEP \textbf{07}, 051 (2004)
doi:10.1088/1126-6708/2004/07/051
[arXiv:hep-th/0404261 [hep-th]].


\bibitem{Henneaux:2006hk}
M.~Henneaux, C.~Martinez, R.~Troncoso and J.~Zanelli,
Annals Phys. \textbf{322}, 824-848 (2007)
doi:10.1016/j.aop.2006.05.002
[arXiv:hep-th/0603185 [hep-th]].



\bibitem{Hamilton}
R.~S.~Hamilton, 
J.\ Diff.\ Geom. {\bf 17}, 255 (1982).
doi:10.4310/jdg/1214436922


\bibitem{Yamabe}
R.~Hamilton, 
``The Ricci flow on surfaces," Mathematics and general relativity (Santa Cruz, CA, 1986). Contemp. Math. Vol. {\bf 71}, 
Amer. Math. Soc., Providence, RI. pp. 237-262 (1988). 
doi:10.1090/conm/071/954419


\bibitem{Perelman:2006un}
G.~Perelman,
[arXiv:math/0211159 [math.DG]].





\bibitem{Roberts:1989sk}
M.~D.~Roberts,
Gen. Rel. Grav. \textbf{21}, 907-939 (1989)
doi:10.1007/BF00769864



















\bibitem{Klemm:2015uba}
D.~Klemm, M.~Nozawa and M.~Rabbiosi,
Class. Quant. Grav. \textbf{32}, no.20, 205008 (2015)
doi:10.1088/0264-9381/32/20/205008
[arXiv:1506.09017 [hep-th]].

\bibitem{Faedo:2015jqa}
F.~Faedo, D.~Klemm and M.~Nozawa,
JHEP \textbf{11}, 045 (2015)
doi:10.1007/JHEP11(2015)045
[arXiv:1505.02986 [hep-th]].

\bibitem{Nozawa:2020gzz}
M.~Nozawa,
Phys. Rev. D \textbf{103}, no.2, 024005 (2021)
doi:10.1103/PhysRevD.103.024005
[arXiv:2010.07561 [gr-qc]].




\bibitem{Lu:2014ida}
H.~Lu and J.~F.~Vazquez-Poritz,
Phys. Rev. D \textbf{91}, no.6, 064004 (2015)
doi:10.1103/PhysRevD.91.064004
[arXiv:1408.3124 [hep-th]].

\bibitem{Lu:2014sza}
H.~L\"u and J.~F.~V\'azquez-Poritz,
JHEP \textbf{12}, 057 (2014)
doi:10.1007/JHEP12(2014)057
[arXiv:1408.6531 [hep-th]].



\bibitem{NT1}
M.~Nozawa and T.~Torii,
Phys. Rev. D \textbf{107}, no.6, 064064 (2023)
doi:10.1103/PhysRevD.107.064064
[arXiv:2211.06517 [hep-th]].





\bibitem{Ellis:1973yv}
H.~G.~Ellis,
J. Math. Phys. \textbf{14}, 104-118 (1973)
doi:10.1063/1.1666161

\bibitem{Bronnikov:1973fh}
K.~A.~Bronnikov,
Acta Phys. Polon. B \textbf{4}, 251-266 (1973)



\bibitem{Martinez:2020hjm}
C.~Martinez and M.~Nozawa,
Phys. Rev. D \textbf{103}, no.2, 024003 (2021)
doi:10.1103/PhysRevD.103.024003
[arXiv:2010.05183 [gr-qc]].

\bibitem{Nozawa:2020wet}
M.~Nozawa,
Phys. Rev. D \textbf{103}, no.2, 024004 (2021)
doi:10.1103/PhysRevD.103.024004
[arXiv:2010.07560 [gr-qc]].










\bibitem{Fan:2016yqv}
Z.~Y.~Fan, B.~Chen and H.~Lu,
JHEP \textbf{05}, 170 (2016)
doi:10.1007/JHEP05(2016)170
[arXiv:1601.07246 [hep-th]].









\bibitem{Misner:1964je}
C.~W.~Misner and D.~H.~Sharp,
Phys. Rev. \textbf{136}, B571-B576 (1964)
doi:10.1103/PhysRev.136.B571



\bibitem{Hayward:1994bu}
S.~A.~Hayward,
Phys. Rev. D \textbf{53}, 1938-1949 (1996)
doi:10.1103/PhysRevD.53.1938
[arXiv:gr-qc/9408002 [gr-qc]].

\bibitem{Maeda:2007uu}
H.~Maeda and M.~Nozawa,
Phys. Rev. D \textbf{77}, 064031 (2008)
doi:10.1103/PhysRevD.77.064031
[arXiv:0709.1199 [hep-th]].


\bibitem{Ohashi:2015xaa}
S.~Ohashi and M.~Nozawa,
Phys. Rev. D \textbf{92}, 064020 (2015)
doi:10.1103/PhysRevD.92.064020
[arXiv:1507.04496 [gr-qc]].

\bibitem{Hawking:1968qt}
S.~Hawking,
J. Math. Phys. \textbf{9}, 598-604 (1968)
doi:10.1063/1.1664615

\bibitem{Hawking:1973uf}
S.~W.~Hawking and G.~F.~R.~Ellis,
Cambridge University Press, 2023,
ISBN 978-1-00-925316-1, 978-1-00-925315-4, 978-0-521-20016-5, 978-0-521-09906-6, 978-0-511-82630-6, 978-0-521-09906-6
doi:10.1017/9781009253161


\bibitem{Hayward:1993wb}
S.~A.~Hayward,
Phys. Rev. D \textbf{49}, 6467-6474 (1994)
doi:10.1103/PhysRevD.49.6467

\bibitem{Nozawa:2007vq}
M.~Nozawa and H.~Maeda,
Class. Quant. Grav. \textbf{25}, 055009 (2008)
doi:10.1088/0264-9381/25/5/055009
[arXiv:0710.2709 [gr-qc]].


\bibitem{Fernandez-Alvarez:2021yog}
F.~Fern\'andez-\'Alvarez and J.~M.~M.~Senovilla,
Class. Quant. Grav. \textbf{39}, no.16, 165012 (2022)
doi:10.1088/1361-6382/ac395b
[arXiv:2105.09167 [gr-qc]].


\bibitem{Bonga:2023eml}
B.~Bonga, C.~Bunster and A.~P\'erez,
Phys. Rev. D \textbf{108}, no.6, 064039 (2023)
doi:10.1103/PhysRevD.108.064039
[arXiv:2306.08029 [gr-qc]].





\bibitem{Maeda:2011sh}
K.~i.~Maeda and M.~Nozawa,
Prog. Theor. Phys. Suppl. \textbf{189}, 310-350 (2011)
doi:10.1143/PTPS.189.310
[arXiv:1104.1849 [hep-th]].

\bibitem{Townsend:2007nm}
P.~K.~Townsend,
J. Phys. A \textbf{41}, 304014 (2008)
doi:10.1088/1751-8113/41/30/304014
[arXiv:0710.5709 [hep-th]].




\bibitem{Headrick:2006ti}
M.~Headrick and T.~Wiseman,
Class. Quant. Grav. \textbf{23}, 6683-6708 (2006)
doi:10.1088/0264-9381/23/23/006
[arXiv:hep-th/0606086 [hep-th]].


\bibitem{Emparan:2019obu}
R.~Emparan and R.~Suzuki,
JHEP \textbf{07}, 094 (2019)
doi:10.1007/JHEP07(2019)094
[arXiv:1905.01062 [hep-th]].












\bibitem{Woolgar:2007vz}
E.~Woolgar,
Can. J. Phys. \textbf{86}, 645-651 (2008)
doi:10.1139/P07-146
[arXiv:0708.2144 [hep-th]].


\bibitem{Polchinski:1998rq}
J.~Polchinski,
Cambridge University Press, 2007,
ISBN 978-0-511-25227-3, 978-0-521-67227-6, 978-0-521-63303-1
doi:10.1017/CBO9780511816079





\bibitem{reviewRF}
B. Kleiner, J. Lott, 
[arXiv:math/0605667 [math.DG]].


\bibitem{deGennes}
P. G. de Gennes, 
C. R. Acad. Sci. Paris II, {\bf 298}, 475 (1984).



\bibitem{King}
J.~R.~King, 
Eur. J. Appl. Math. {\bf 5}, 359, (1994).
doi:10.1017/S0956792500001509 


\bibitem{Rosenau}
P. Rosenau, 
Phys. Rev. Lett. {\bf 74}, 1056, (1995).
doi:10.1103/PhysRevLett.74.1056





\bibitem{Caldarelli:2012hy}
M.~M.~Caldarelli, J.~Camps, B.~Gout\'eraux and K.~Skenderis,
Phys. Rev. D \textbf{87}, no.6, 061502 (2013)
doi:10.1103/PhysRevD.87.061502
[arXiv:1211.2815 [hep-th]].

\bibitem{Caldarelli:2013aaa}
M.~M.~Caldarelli, J.~Camps, B.~Gout\'eraux and K.~Skenderis,
JHEP \textbf{04}, 071 (2014)
doi:10.1007/JHEP04(2014)071
[arXiv:1312.7874 [hep-th]].

\bibitem{Wu:2022gpm}
T.~Wu,
Phys. Rev. D \textbf{108}, no.4, 044001 (2023)
doi:10.1103/PhysRevD.108.044001
[arXiv:2209.02278 [gr-qc]].



\bibitem{Maeda:2015cia}
H.~Maeda,
Class. Quant. Grav. \textbf{32}, no.13, 135025 (2015)
doi:10.1088/0264-9381/32/13/135025
[arXiv:1501.03524 [gr-qc]].








\end{thebibliography}
\end{document}